\documentclass[a4paper,11pt]{article}
\pdfoutput=1 

\usepackage{jheppub} 

\usepackage[T1]{fontenc} 
\usepackage{amsmath}
\usepackage{amssymb}
\usepackage{slashed}
\usepackage{color}

\DeclareRobustCommand{\Sec}[1]{Sec.~\ref{#1}}

\DeclareRobustCommand{\App}[1]{App.~\ref{#1}}
\DeclareRobustCommand{\Tab}[1]{Table~\ref{#1}}

\DeclareRobustCommand{\Fig}[1]{Fig.~\ref{#1}}
\DeclareRobustCommand{\Figs}[2]{Figs.~\ref{#1} and \ref{#2}}
\DeclareRobustCommand{\Eq}[1]{Eq.~(\ref{#1})}
\DeclareRobustCommand{\Eqs}[2]{Eqs.~(\ref{#1}), (\ref{#2})}

\definecolor{red1}{cmyk}{0,1,1,0.3}

\newcommand{\cO}{\mathcal{O}}
\newcommand{\cL}{\mathcal{L}}

\newcommand{\cR}{\mathcal{R}}

\newcommand{\TeV}{\mathrm{TeV}}
\newcommand{\GeV}{\mathrm{GeV}}
\newcommand{\MeV}{\mathrm{MeV}}

\newcommand{\fb}{\mathrm{fb}}
\newcommand{\ab}{\mathrm{ab}}

\newcommand{\mm}{\mathrm{mm}}
\newcommand{\cm}{\mathrm{cm}}

\newcommand{\eg}{\textit{e.g.}}
\newcommand{\ie}{\textit{i.e.}}
\newcommand{\sigang}{\sigma_{\text{ang}}}
\newcommand{\sigpos}{\sigma_{\text{pos}}}

\newcommand{\FASERnu}{FASER$\nu$}
\newcommand{\FASERnuTwo}{FASER$\nu$2}
\newcommand{\DFASERnu}{\Delta_{\text{\FASERnu}}}
\newcommand{\DFASERnuTwo}{\Delta_{\text{\FASERnuTwo}}}
\newcommand{\SFasernu}{S_{\text{\FASERnu}}}
\newcommand{\SFasernuTwo}{S_{\text{\FASERnuTwo}}}
\newcommand{\LLHC}{\mathcal{L}_{\text{\tiny LHC}}}
\newcommand{\LLHCHL}{\mathcal{L}_{\text{\tiny HL-LHC}}}

\newcommand{\Madgraph}{\textsc{madgraph}}

\title{Hunting muonic forces at emulsion detectors}

\author[a,b]{Akitaka Ariga,}
\author[c]{Reuven Balkin,}
\author[c]{Iftah Galon,}
\author[c]{Enrique Kajomovitz,}
\author[c]{Yotam Soreq}

\affiliation[a]{Albert Einstein Center for Fundamental Physics, Laboratory for High Energy Physics, University of Bern, Sidlerstrasse 5, CH-3012 Bern, Switzerland}
\affiliation[b]{Department of Physics, Chiba University, 1-33 Yayoi-cho Inage-ku, 263-8522 Chiba, Japan}
\affiliation[c]{Physics Department, Technion -- Israel Institute of Technology, Haifa 3200003, Israel}

\emailAdd{reuven.b@campus.technion.ac.il}
\emailAdd{iftah.galon@gmail.com}
\emailAdd{enrique@physics.technion.ac.il}
\emailAdd{akitaka.ariga@lhep.unibe.ch}
\emailAdd{soreqy@physics.technion.ac.il}

\abstract{
Only two types of Standard Model particles are able to propagate the $480\,$meters separating the ATLAS interaction point and FASER: neutrinos and muons. 
Furthermore, muons are copiously produced in proton collisions. 
We propose to use \FASERnu{} as a muon fixed target experiment in order to search for new bosonic degrees of freedom coupled predominantly to muons. 
These muon force carriers are particularly interesting in light of the recent measurement of the muon anomalous magnetic moment. 
Using a novel analysis technique, we show that even in the current LHC run, \FASERnu{} could potentially probe previously unexplored parts of the parameter space. 
In the high-luminosity phase of the LHC, we find that the improved sensitivity of \FASERnuTwo{} will probe unexplored parameter space and may be competitive with dedicated search proposals.}

\begin{document} 

\maketitle
\flushbottom

\section{Introduction}
\label{sec:intro}

The Standard Model of particle physics~(SM) successfully describes Nature in a wide range of energy scales. 
However, there is experimental evidence and strong theoretical arguments for the existence of new physics~(NP) beyond the SM~(BSM), see \eg~\cite{EuropeanStrategyforParticlePhysicsPreparatoryGroup:2019qin}. 
New feebly interacting particles~(FIPs) at the MeV--to--GeV mass range are well-motivated in many BSM scenarios and have recently received a lot of attention, both on the theoretical and experimental side~\cite{Agrawal:2021dbo,Lanfranchi:2020crw}.
One particular subset of FIPs are bosons with non-universal couplings to the SM fermions, which appear in various extensions of the SM~\cite{Batell:2016ove,Chen:2017awl,Batell:2017kty,Marsicano:2018vin}.

Bosons which couple predominantly to muons, often referred to as muonic force carriers~(MFCs), are motivated by the recent measurement of the muon anomalous magnetic moment, $(g-2)_\mu$~\cite{Bennett_2006,PhysRevLett.126.141801}. 
Comparing the experimental data to the data-driven SM prediction points to an anomaly, which can be explained by a weakly-coupled sub GeV MFC \eg~\cite{Pospelov:2008zw,Jegerlehner:2009ry}. 
However, recent lattice results~\cite{Borsanyi:2020mff,Alexandrou:2022amy,Ce:2022kxy} seem to suggest the data is in fact consistent with the SM. 
Moreover, a recent measurement of the $e^+e^-\to \pi\pi$ cross section even appears to alleviate the tension between the data-driven SM prediction and the experimental result~\cite{CMD-3:2023alj}.

On the other hand, it is possible that NP contributions to $(g-2)_\mu$ are suppressed due to cancellation at the quantum level. 
This was demonstrated in~\cite{Balkin:2021rvh}, where an approximate cancellation at the 1-loop level arises due to a global symmetry. 
This type of scenario provides a clear motivation for more direct, tree-level searches, such as the one proposed in this work.

Beyond the precision measurement of $(g-2)_\mu$, MFCs were targeted by direct searches in flavor factories such as BaBar~\cite{BaBar:2016sci}, Belle II~\cite{Belle-II:2019qfb} and NA62~\cite{Krnjaic:2019rsv,NA62:2021bji}, while vector MFCs were also probed by neutrino beam experiments~\cite{CHARM-II:1990dvf,CCFR:1991lpl,Altmannshofer_2014}.
In addition, there is a bound from supernova SN1987~\cite{Croon:2020lrf}.
Proposed experiments such as NA64$_\mu$~\cite{Chen:2017awl,Chen:2018vkr,Gninenko:2019qiv}, M$^3$~\cite{Kahn:2018cqs} and proton beam dump~\cite{Forbes:2022bvo} can potentially probe unexplored regions of the MFC parameter space. 
An alternative approach is to maximize the NP reach of running experiments.
For example, it was shown that the ALTAS experiment can probe the MFC parameter space via missing momentum measurements~\cite{Galon:2019owl}.

In this work, we present a novel method to search for MFCs using emulsion detectors as muon fixed targets.
In particular, we estimate the sensitivity of \FASERnu~\cite{FASER:2019dxq,FASER:2020gpr,FASER:2021mtu}, the front part of the running Forward Search Experiment~(FASER)~\cite{Feng_2018,FASER:2018ceo,FASER:2018bac,FASER:2018eoc}, to the MFC parameter space.
Neutrinos and muons are the only SM particles able to propagate the $480$\,meters separating the ATLAS interaction point and FASER. 
The muons are produced in proton collisions, mostly as decay products or in secondary interactions between the collision products and the surrounding matter.
Although most muons are deflected, an estimate of $N_\mu \sim 10^9$ muons with energy above $100\,\GeV$ are expected to reach \FASERnu{} with $150\,\fb^{-1}$ integrated luminosity~\cite{FASER:2018eoc}. 
Using missing energy signatures, we estimate that MFC couplings as small as $10^{-3}-10^{-4}$ may be within reach in \FASERnu{} and even smaller with \FASERnuTwo, which is expected to be installed at the LHC high luminosity~(HL-LHC) stage. 
The sensitivity crucially depends on the background rejection efficiency, which must be carefully estimated in a dedicated detector study left for future work.

\section{Simplified model}
\label{sec:model}

We consider a simplified model in which the muon is coupled to a light mediator $\chi=\{S,V\}$, with the following interaction Lagrangians,
\begin{align}
    \label{eq::L_int}
    \cL_S 
=   g_S \,S \,\bar{\mu} \mu \,  \qquad\text{and}\qquad  
    \cL_V = g_V V_\alpha \,\bar{\mu} \gamma^\alpha\mu \,,
\end{align}
for a scalar and vector MFC, respectively, with the masses denoted by $m_\chi$.  
Pseudo scalar and axial vector interactions are expected to give similar results.
The effective theory described by \Eq{eq::L_int} can emerge as a low energy limit of a UV complete theory;  for example scalar interactions can arise as a result of integrating out heavy leptons~\cite{Krnjaic:2019rsv}, while the vector interactions naturally arise in a spontaneously broken gauged $U(1)_{L_\mu-L_\tau}$ theory ~\cite{He:1990pn,Foot:1990mn,PhysRevD.44.2118}. 

Assuming $\chi$ is negligibly coupled to other SM constituents (\eg{} the irreducible one-loop level coupling of $S$ to photons), we consider the case where it is either long-lived at the detector scale or decays invisibly. 
This scenario can arise naturally in the muonphillic case when $m_\chi < 2m_\mu$, while above the di-muon threshold some more assumptions are required \eg{} sizable decay rate to a dark sector.
Either way, once scattered against a fixed-target $N$, some muons lose a fraction of their energy due to the 2-to-3 process,
\begin{align}
    \label{eq::fixed_target_process}
    N+\mu \to N+\mu+\chi\,,
\end{align}
while $\chi$ or its dark sector decay products escape the system without depositing any energy in the detector. 
Thus, $\chi$ can be searched for using missing-momentum signatures in \FASERnu, utilizing it as a muon-fixed-target experiment.

\section{Method}
\label{sec:method}

\FASERnu{} is an emulsion detector mounted in front of the FASER main detector.
It is composed of 730 emulsion layers interleaved with 1.1\,mm thick tungsten plates of area $ \SFasernu = 25\,\cm \times 30\,\cm$~\cite{FASER:2023zcr}. 
As muons pass through \FASERnu, their track are measured by the emulsion layers.
Throughout this work, we assume each layer provides only the position of the muon as it passes. 
In principle, due to the finite thickness of the emulsion layer and its structure, additional angular information can be deduced, which can potentially be harnessed to improve the analysis. 
We characterize each muon passing through \FASERnu{} by two simple properties: 
(i)~the ratio between the final and the initial muon energies, $\cR_{fi}\equiv (E_{\mu})_f/(E_{\mu})_i$, and (ii)~the largest scattering angle of the track, $\theta_\mu$, which characterizes the largest kink in the track. 
Large kinks are strongly correlated with large energy losses.   
\begin{figure}[t]
\centering
\includegraphics[width=0.9\textwidth]{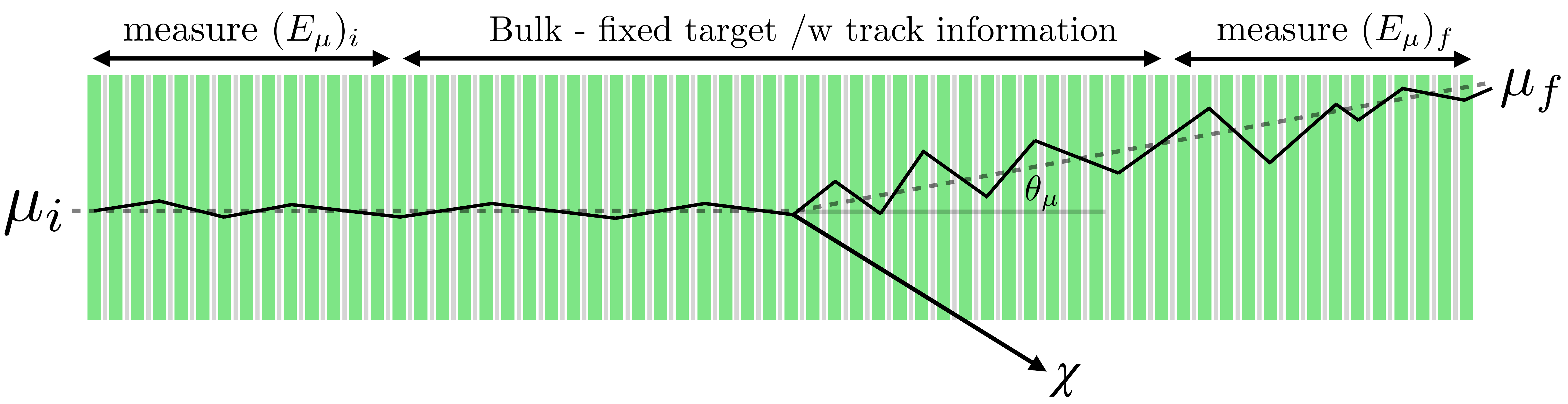}
\caption{A sketch of a muon track at \FASERnu. The initial and final energies can be estimated using the MCS method (see \Sec{sec:EnergyMeas}), where we emphasize using the solid black lines that the MCS angles would generically increase after the muon losses energy. In practice, the MCS angles are $\mathcal{O}(1\,\text{mrad})$ and the muon tracks appear as straight lines (dashed).
}
\label{fig::track}
\end{figure}
To fully benefit from the layered structure of \FASERnu, we propose the detector
acts as an instrumented target:
\begin{itemize} 
    \item The bulk of the detector supplies the target mass and is used as a fixed target for the incoming muon flux.
    \item Using the precise information about the positions of the muon as it propagates through the detector, the front and rear parts of \FASERnu{} are used to measure the incoming and outgoing energies of the muon, respectively, using a method based on multiple Coulomb scattering~(MCS), such as the one outlined in \Sec{sec:EnergyMeas}.
    \item The same spatial information can be used in the bulk of the detector in order to identify the position and magnitude of kinks in the track.
\end{itemize}
Our track characterization is summarized in a sketch shown in \Fig{fig::track}.

\subsection{Muon energy reconstruction}
\label{sec:EnergyMeas}

Let us briefly review an MCS-based method for measuring a particle energy in an emulsion detector~\cite{Kodama:2002dk,OPERA:2011aa,FASER:2019dxq}.
As charged particles pass through medium, they undergo a large number of elastic, small-angle Coulomb scatterings off nuclei. 
Due to the central limit theorem, the cumulative effects of the scatterings are normally distributed~\cite{Bethe:1953va,Scott:1963xw,Motz:1964gpe}. 
The MCS angle $\theta$ of an outgoing, unit-charge particle (relative to the incoming direction) with energy $E \gg m$ is approximately described by a Rayleigh distribution $\sim ({\theta}/{\sigma^2_\theta})e^{-\theta^2/(2\sigma^2_\theta)}$ with~\cite{Lynch:1990sq}
\begin{align}
    \sigma_\theta(X,E)
    &\approx (0.136\,\text{mrad}) \left(\frac{100\,\text{GeV}}{E}\right) \, 
    \sqrt{{X}/{X_0}}\left[ 1+0.038 \log 
    \left( X/X_0 \right) \right]\,,
    \label{eq::sigma_theta}
\end{align}
where $X$ is the in-medium propagation distance and $X_0$ is the material radiation length. 
Thus, the typical single-layer, $X\sim 1.1\,\mm$, MCS angle in \FASERnu{} for a 100\,GeV muon is roughly $\sigma_\theta \sim 0.07\,$mrad, where we used $X_0^W = 3.504\,\mm$~\cite{Workman:2022ynf}.
This is to be compared with the single-layer angular resolution in \FASERnu{} of $\sigang \sim 0.23$~mrad, where we used $\sigang = \sqrt{2} \sigpos/X$~\cite{FASER:2019dxq}, with $\sigpos=0.18\times 10^{-3}\,\mm$~\cite{AkiTalk} the spatial resolution of the emulsion plate.

In practice, the muon energy is estimated by measuring linear displacements rather than angles, which are defined as follows. 
Consider three points along the track, denoted by $(x_i,y_i,z_i)$ with $i=0,1,2$. The $z_i$'s are locations of emulsion plates such that $z_2-z_1= z_1-z_0\approx X$ (neglecting the width emulsion layer, to be reintroduced in \Eq{eq::X_scat}). 
The linear displacement in the $x$ coordinate is defined as 
\begin{align}
    s_x \equiv (x_2-x_1)-(x_1-x_0) \, ,    
\end{align}
and similarly for the $y$ coordinate. 
Clearly $s_x=0\,(s_y=0)$ if the three points are on a line in the $x-z\,(y-z)$ plane. 
The variables $s_x$ and $s_y$ are independent random variables, which are normally distributed with the following standard deviation
\begin{align}
    \label{eq::sigma_s}
    \sigma_s(X,E)  
=   \sqrt{\left[\sqrt{2/3}X \, \sigma_\theta(X^\text{scat},E)\right]^2
    +\left[\sqrt{6}\,\sigpos\right]^2}\,,
\end{align}
where
\begin{align}
    \label{eq::X_scat}
    X^\text{scat}(X) 
    \equiv   
    \left(\frac{\Delta_W}{\Delta_W+\Delta_{\text{\tiny emu.}}} \right) X\,,
\end{align}
is the scattering length as a function of the physical propagation length $X$. 
The prefactor takes into account the fact that the finite-width emulsion layer is not a scattering medium.
It is given by the ratio of the length of the scattering medium, i.e. $\Delta_W=1.1\,\mm$ of tungsten, and the physical propagation distance $\Delta_W+\Delta_{\text{\tiny emu.}}$ , where
$\Delta_{\text{\tiny emu.}}=0.34\,\mm$ is the width of the emulsion plate. 
For a given track, \Eq{eq::sigma_s} can be calculated for various values of $X$. 
The energy $E$ can then be inferred by performing a fit to the data. 

The FASER collaboration reported an energy resolutions of $46\%$ and $57\%$ for $E=200\,\GeV$ and $1\,\TeV$, respectively, applying the MCS method using $100$ layers~\cite{FASER:2019dxq}. 
We analyzed simulated data at lower energies, which yielded similar results, for more details see \App{app:energy_meas}. 
By using a larger sample size of 400 (200 layers of independent $x,y$ displacements), the statistics-dominated energy uncertainty is reduced by about a factor of $\sim\sqrt{400/100}\sim2$, namely $\Delta E/E \sim 20\%$. 

An MCS-based method can also be used to determine $\theta_\mu$.
First, we identify the region in the detector in which the muon lost its energy using a sliding window algorithm.
The largest scattering angle measured in this region is strongly correlated with $\theta_\mu$, if the scattering angle is larger than the typical MCS angle and is above the angular resolution. 
This is usually the case for large energy losses, which as we discuss below, is the kinematic region of interest. 
In a preliminary study, we find that even a simple realization of this algorithm leads to a small reduction in signal efficiency, which can be made even smaller with optimization, \eg{} by using machine learning. 
A detailed study is left for future work.  

\subsection{Signal}
\label{sec:signal}

\begin{figure}[t]
\centering
\includegraphics[width=0.45\textwidth]{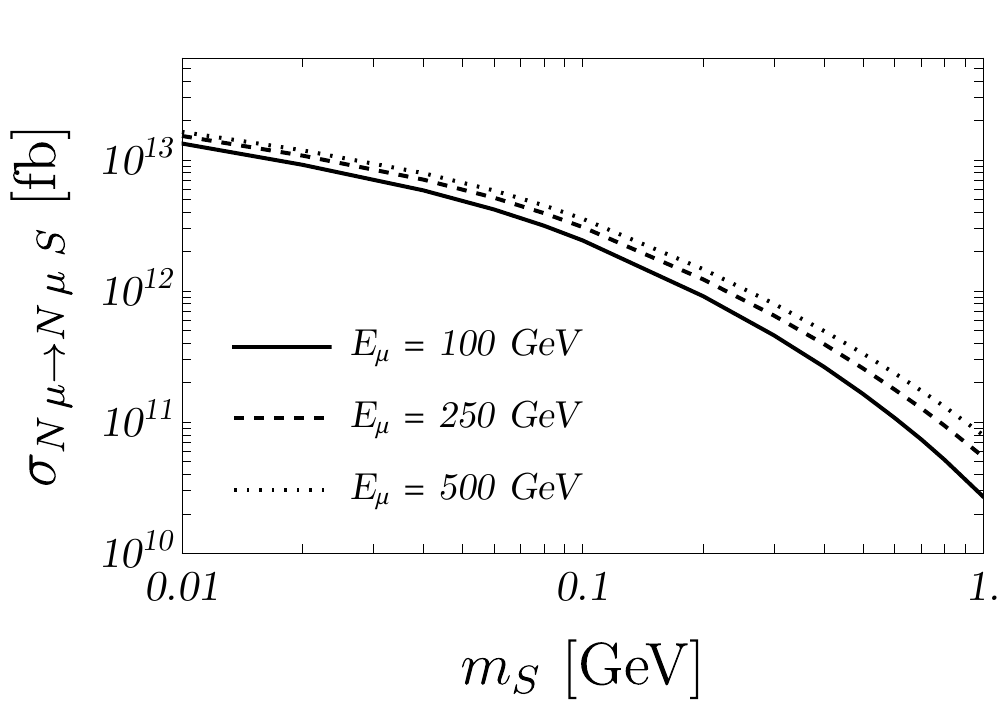}
\includegraphics[width=0.45\textwidth]{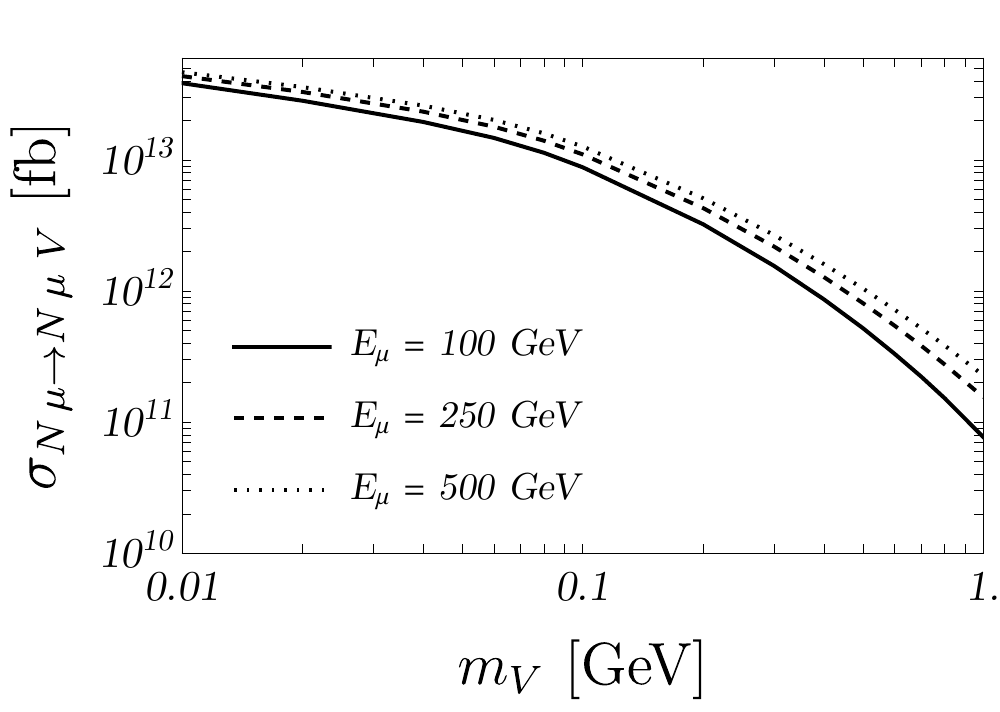}
\caption{The cross-section of the process $N \mu \to N \mu \,\chi$ as a function of $m_\chi$ for $N=\text{W}$ and $\chi=S\,(V)$ left\,(right). 
Both cross section are plotted for $g_S=g_V=1$.}
\label{fig::signal_prod}
\end{figure}

\begin{figure}[t]
\centering
\includegraphics[width=0.9\textwidth]{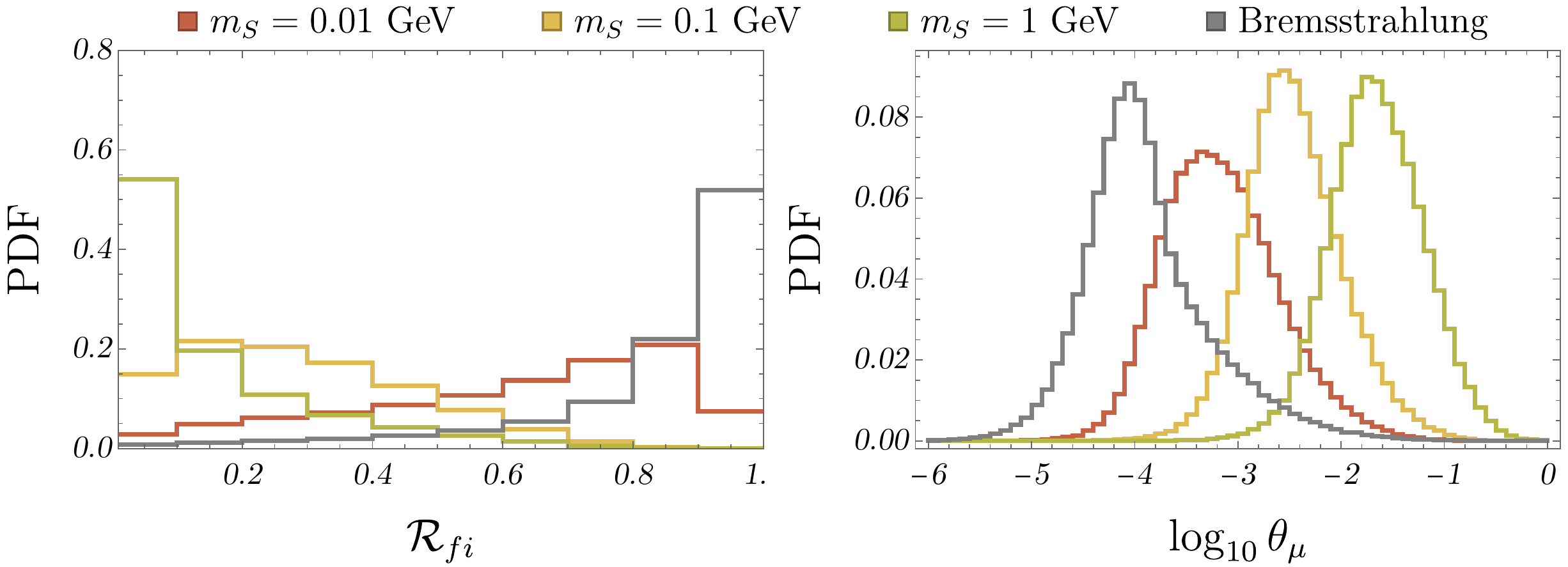}
\\
\includegraphics[width=0.9\textwidth]{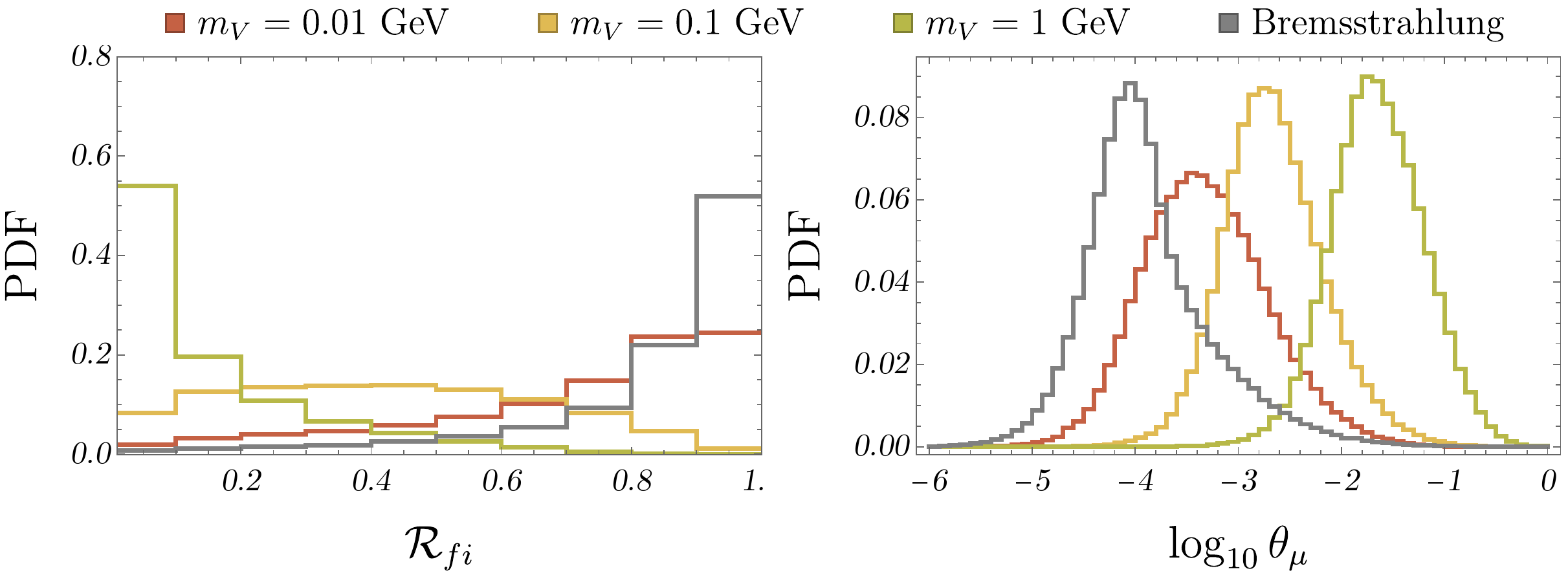}
\caption{Signal $\cR_{fi}$ and $\theta_\mu $ distributions for $(E_\mu)_i=100$~GeV and three representative scalar (top) and vector (bottom) MFC masses, 0.01~GeV, 0.1~GeV and 1~GeV in red, yellow and green, respectively. For comparison, we plot the same distribution for our dominant background due to bremsstrahlung.  }
\label{fig::EfEi_angle_dist}
\end{figure}

\begin{figure}[t]
\centering
\includegraphics[width=0.32\textwidth]{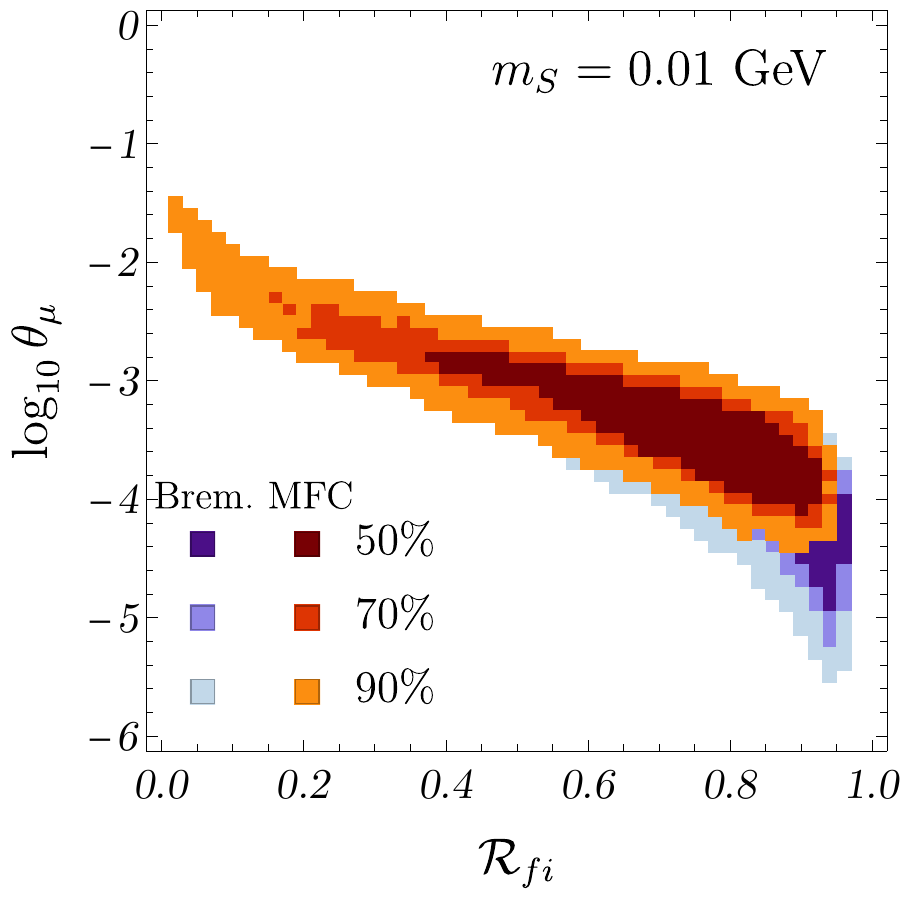}
\includegraphics[width=0.32\textwidth]{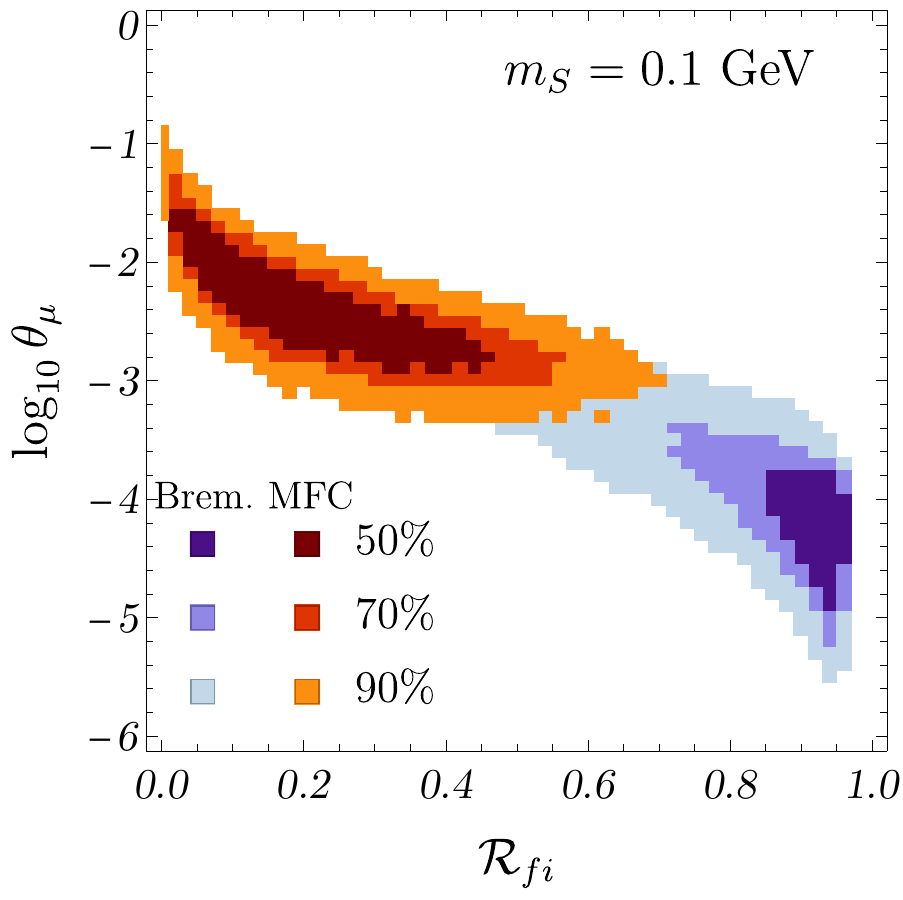}
\includegraphics[width=0.32\textwidth]{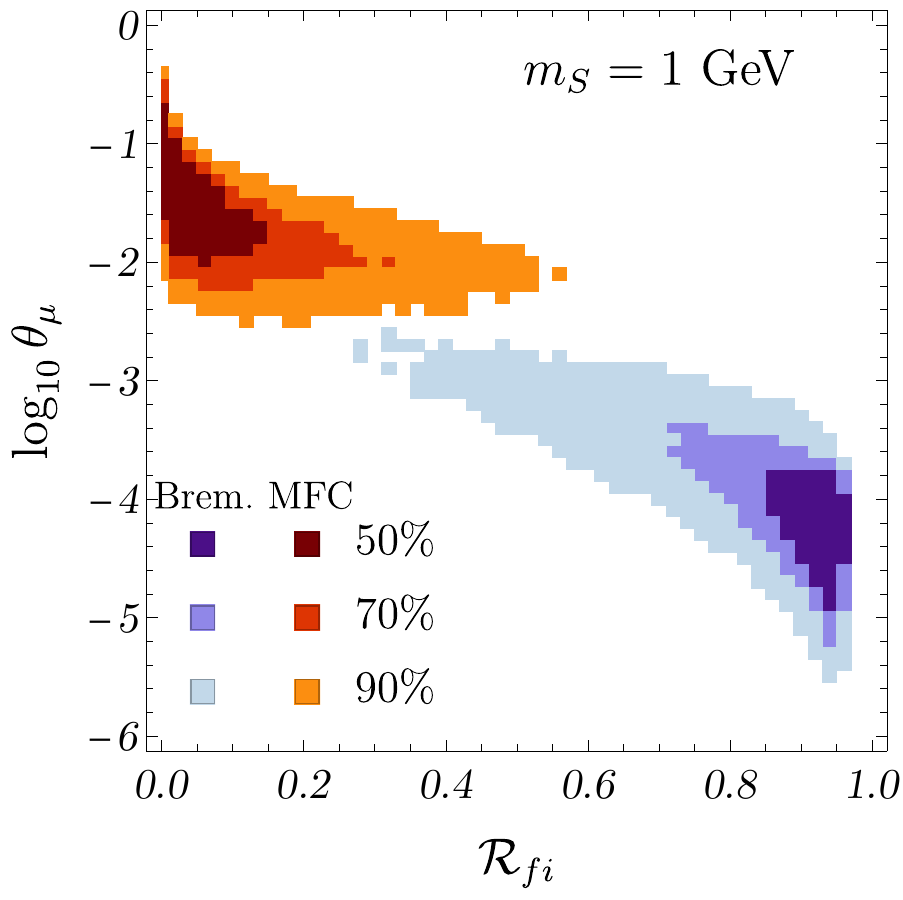}
\\
\includegraphics[width=0.32\textwidth]{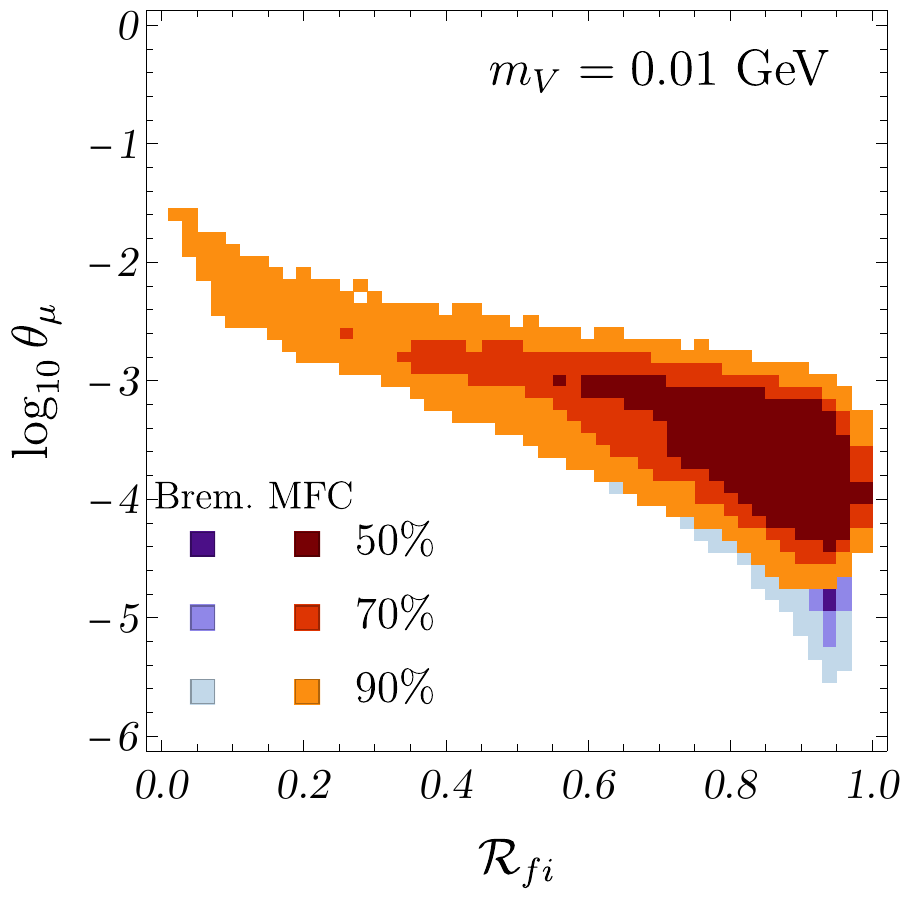}
\includegraphics[width=0.32\textwidth]{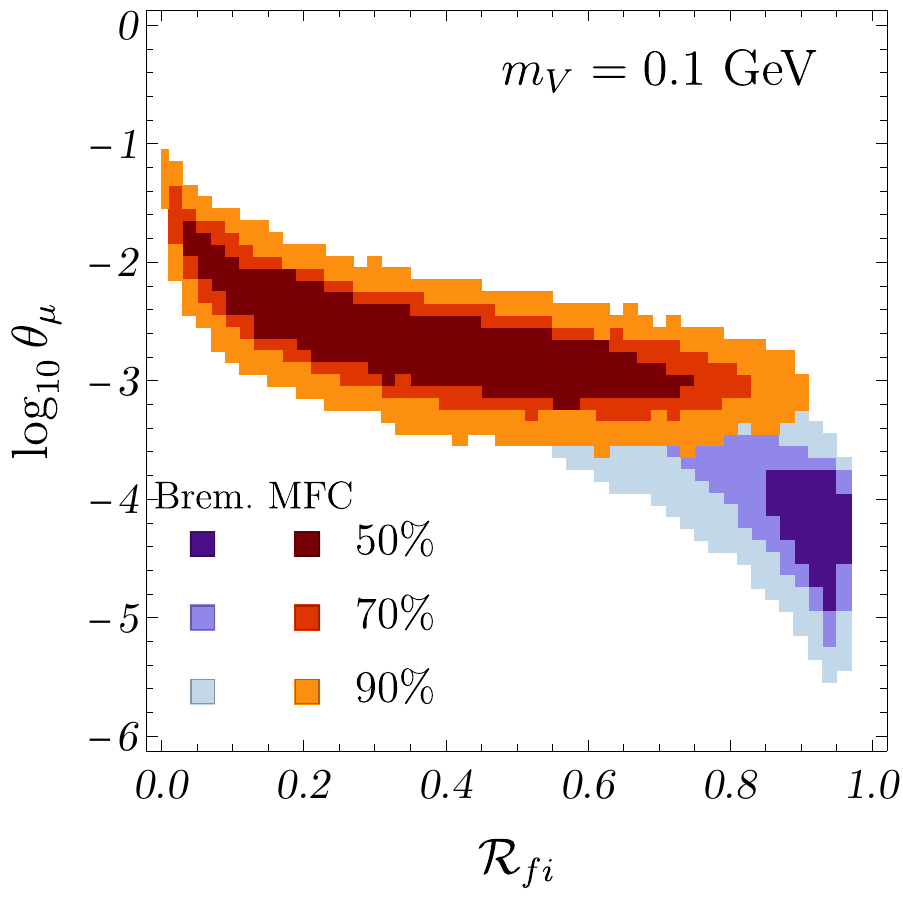}
\includegraphics[width=0.32\textwidth]{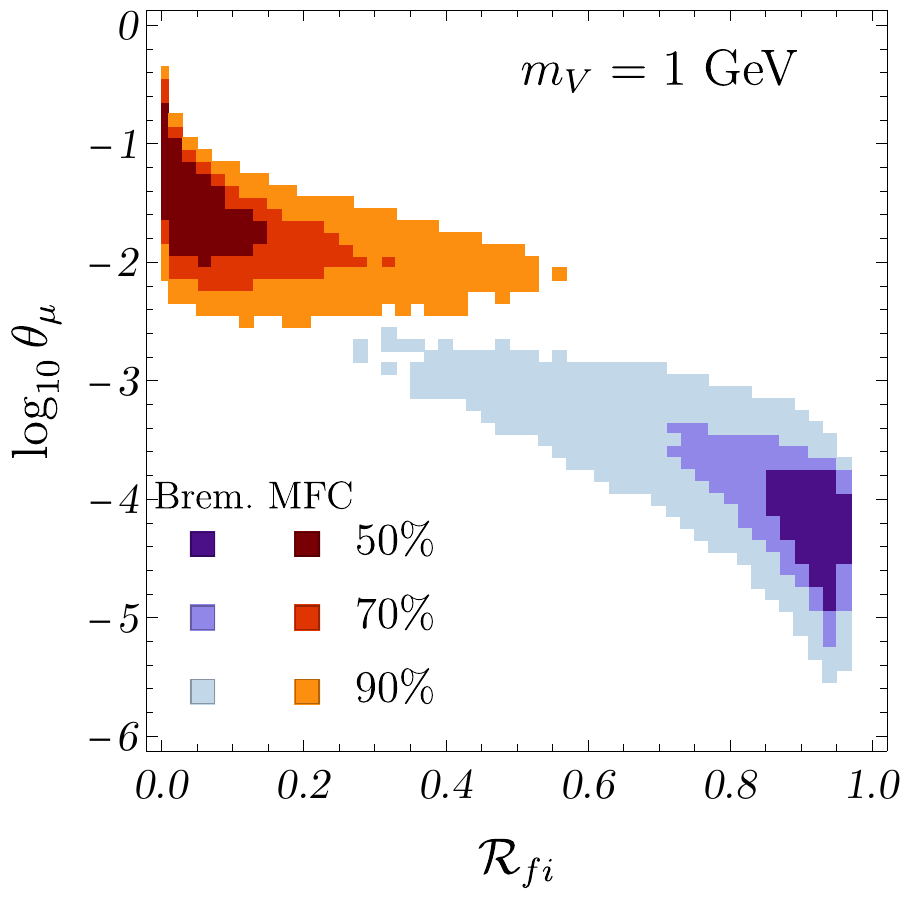}
\caption{2D probability density distribution $\{\cR_{fi},\theta_\mu\} $ for $(E_\mu)_i=100$~GeV and three representative scalar (top) and vector (bottom) MFC masses, 0.01~GeV (left), 0.1~GeV (center) and 1~GeV (right). For comparison, we plot the same distribution for our dominant background due to bremsstrahlung.}
\label{fig::EfEi_angle_2d}
\end{figure}

We calculated the cross-section for the coherent process of \Eq{eq::fixed_target_process} (taking $N$ to be tungsten atoms) numerically using \Madgraph~\cite{Alwall:2014hca} for a range of MFC masses and initial muon energies, see \Fig{fig::signal_prod}. 
The cross section depends only weakly on the initial energy of the muon and decreases as expected for heavier MFC masses.
Each MFC emission event can be characterized by $\cR_{fi}$ and $\theta_\mu$. 
As shown in~\Fig{fig::EfEi_angle_dist}, heavier MFC masses typically lead to larger scattering angles and energy loss, which are strongly correlated, see \Fig{fig::EfEi_angle_2d}.
Lighter MFC masses, on the other hand, lead to smaller scattering angles and energy loss and are increasingly SM-like.  
For comparison, we also plot the corresponding distributions from the main background due to bremsstrahlung, to be discussed below in \Sec{sec:background}. 


\subsection{Background}
\label{sec:background}

The propagation of muons through matter has been studied in detail, see \eg~\cite{Workman:2022ynf}. 
The main SM processes which contribute to the energy loss of the muon as it propagates through matter are, by probability order at small energy losses, 
\begin{align}
    \label{eq::BG_proc}
    N+\mu \to 
    \begin{cases}
    N + \mu + e^-+e^+ &\text{(Pair production)} \\
    N^+ + \mu + e^-\;\;\;\;&\text{(Ionization)} \\
    N + \mu + \gamma &\text{(Bremsstrahlung)}   \\
    N^* + \mu + ... &\text{(Nuclear)}
    \end{cases}\,.
\end{align} 
As the muon propagates through the tungsten layers, its total energy loss is typically an accumulation of a large number of scatterings due to the processes of \Eq{eq::BG_proc}. 
At $100\,$GeV, the average energy loss in tungsten is $\langle dE/dx \rangle = 3.05\,\MeV \cm^2/\text{gr}$~\cite{Workman:2022ynf}. 
Thus, the average energy loss at \FASERnu{} would be $\langle \Delta E \rangle = 4.97\,\GeV$ per muon. 

\begin{figure}[t]
\centering
\includegraphics[width=0.95\textwidth]{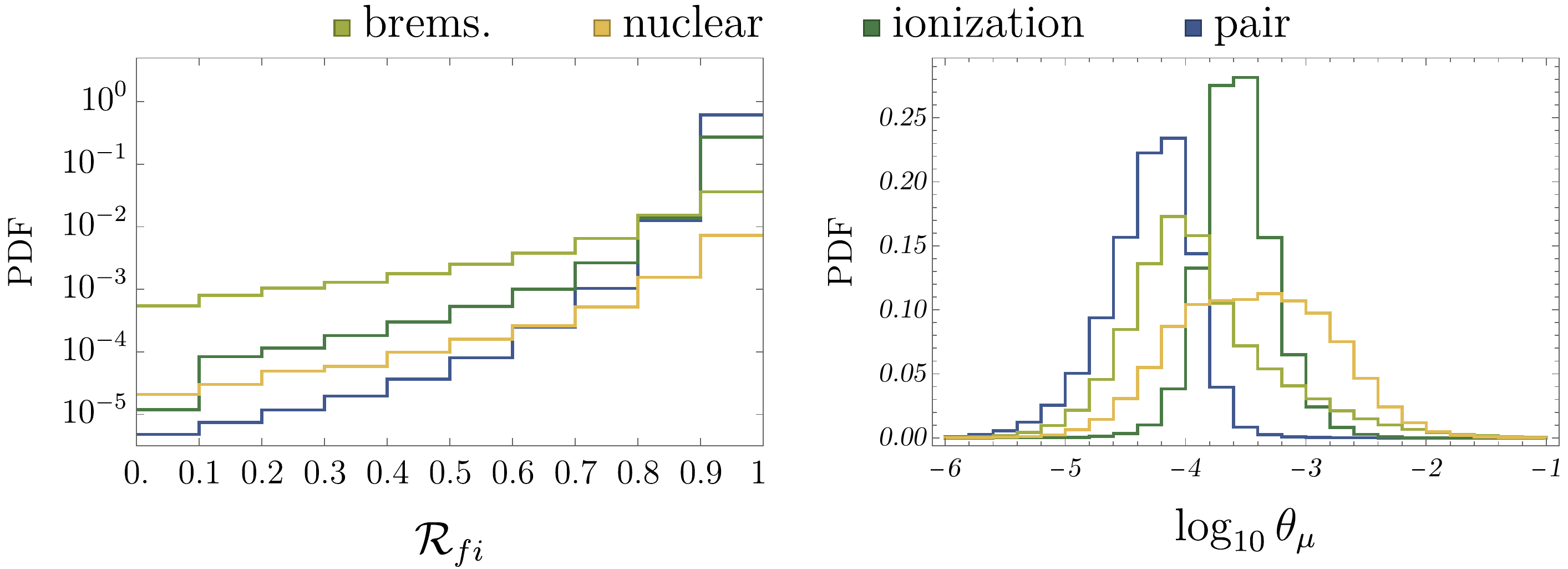}
\caption{Background $\cR_{fi}$ (left) and $\theta_\mu $ (right) distributions for bremsstrahlung (light green), nuclear (yellow), ionization (dark green) and pair production (blue) processes of \Eq{eq::BG_proc} with $(E_\mu)_i=100\,\GeV$. Note that the $\cR_{fi}$ distributions are weighted according to the values in \Tab{tab::SM_proc}, such that the sum of all four curves is normalized to 1.}
\label{fig::SM_dist}
\end{figure}

\begin{figure}[t]
\centering
\includegraphics[width=0.45\textwidth]{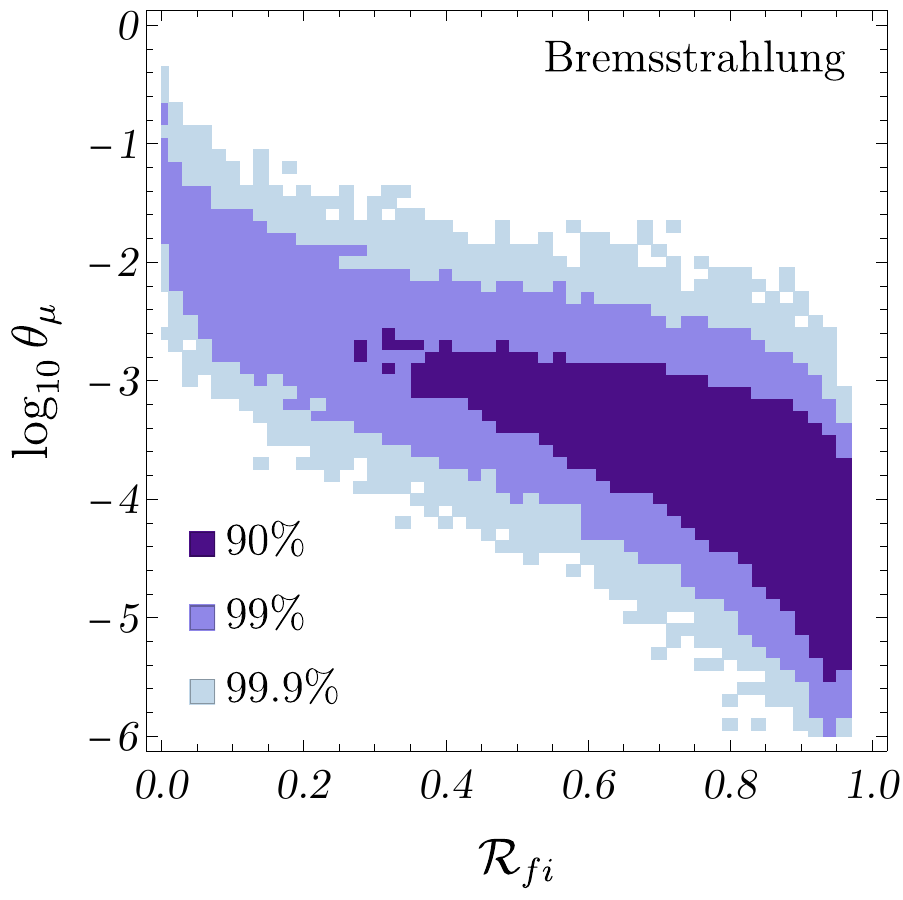}
\includegraphics[width=0.45\textwidth]{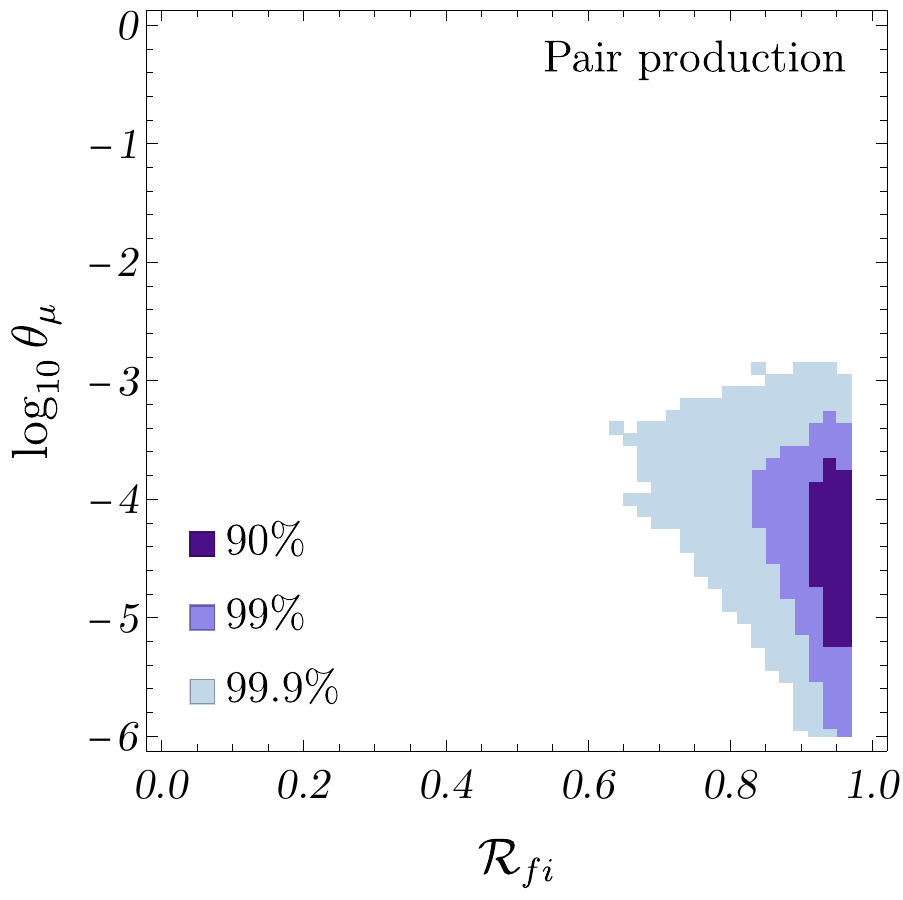}
\\
\includegraphics[width=0.45\textwidth]{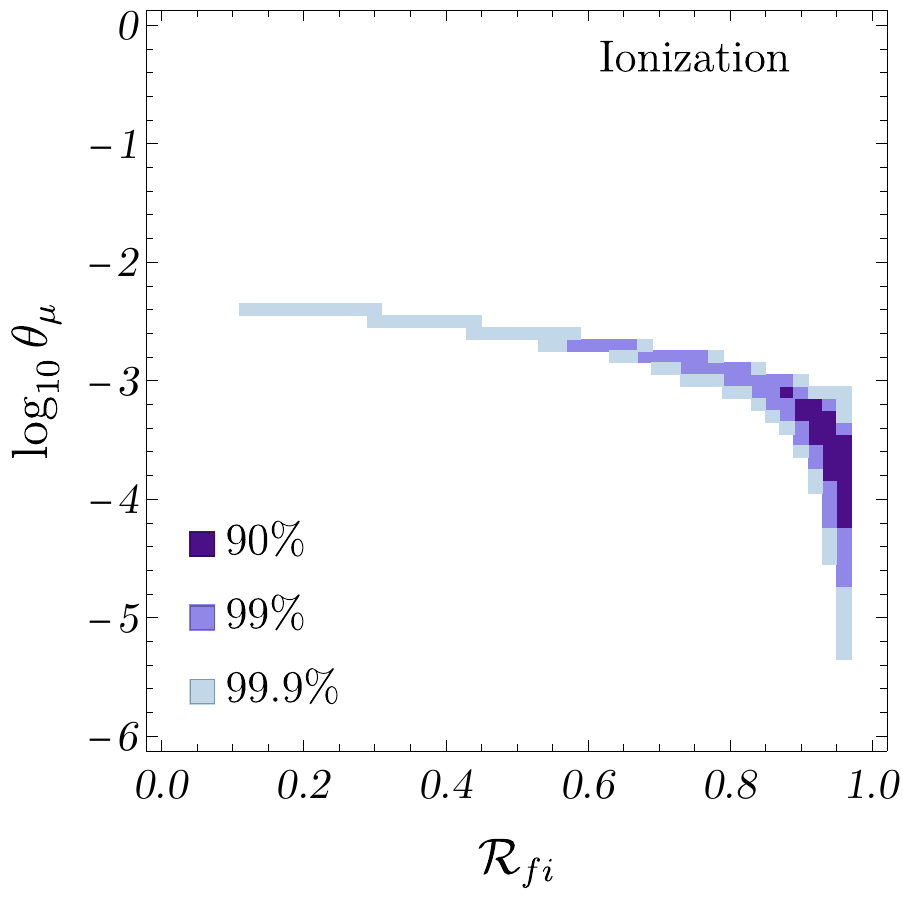}
\includegraphics[width=0.45\textwidth]{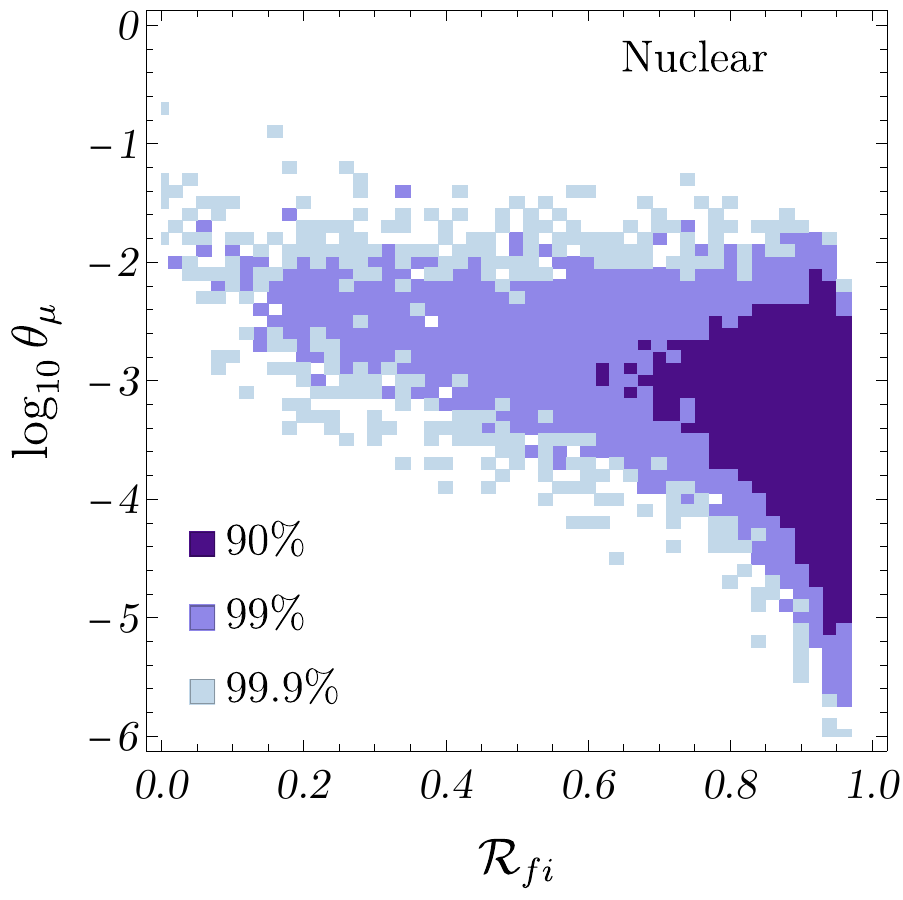}
\caption{2D probability density distribution $\{\cR_{fi},\theta_\mu\} $ for bremsstrahlung, nuclear, ionization and pair production process shown in \Eq{eq::BG_proc} with $(E_\mu)_i=100$~GeV. }
\label{fig::SM_EfEi_angle_2d}
\end{figure}

Our main source of background are rare events in which the muon loses a large fraction of its energy due to a SM process. 
In order to study the background processes, we simulated $\sim 5\times 10^6$ muon tracks as they propagate through a simplified model of \FASERnu\footnote{The tungsten layers are modeled accurately, but the film is treated as single measurement per film layer.}
using GEANT4~\cite{GEANT4:2002zbu,Bogdanov:2006kr} with various values for the initial energy.
We associate each of the $\sim 5\times 10^6$ simulated tracks with one of the SM processes listed in \Eq{eq::BG_proc} according to its single largest energy loss event during its propagation. 
This identification is useful for tracks in which a rare hard scattering occurred, while tracks with small energy losses, as explained above, are typically a result of many soft scatterings. 
We present the distribution of tracks in \Tab{tab::SM_proc}. 
The $\cR_{fi}$ and $\theta_\mu$ probability distributions for each individual process are plotted in \Fig{fig::SM_dist}. 
The correlation between the kinematic variables can be seen in the density plots of \Fig{fig::SM_EfEi_angle_2d}. 
As expected, the vast majority of these scattering events generate small energy loss, $\Delta E \lesssim \langle \Delta E \rangle$, while the rare large energy loss events are dominated by bremsstrahlung.  

\begin{table}[t]
\centering
\begin{tabular}{ |c|c|c|c|c|c| }
\hline
&Pair & Ionization & Brems. & Nuclear& Total  \\ 
 \hline  \hline
 \# of tracks &$3.15\times 10^6$& $1.42\times 10^6$ & $3.45\times 10^5$ & $4.7\times 10^4$ &$4.96\times 10^6$
 \\   \hline
  Fraction  & $0.63$ & $0.29$ & $0.07$ & $0.009$ & 1.0
 \\ \hline
\end{tabular}
\caption{Distribution of tracks according to the SM process responsible for the largest energy loss along the track. This distribution was calculated for initial muon energy of $ 100\,\GeV$. Note that at these high energies, the dependence of these rates on the initial muon energy is negligible.}
\label{tab::SM_proc}
\end{table}


The background~(BG) rejection for the three prompt SM processes, \ie{} ionization, pair production and nuclear, quantifies the ability to detect energy depositions in the close vicinity of a given muon track. 
This can be done either by (1) finding a statistically significant excess of secondary particles around the muon track, or by (2) associating an emerging high-energy particle track to the muon track. 
The transverse length scale of EM showers comprised from secondary particles is given by the Moli\`ere radius, $r_{\text{\tiny M}}=9.3\,\mm$ in tungsten~\cite{Workman:2022ynf}. 
The other relevant scale is the average inter-track distance $r_{\text{track}}\sim \rho^{-1/2}$, where $\rho$ is the average track surface density. 
If $r_{\text{track}} \gg r_{\text{\tiny M}}$, the tracks are sufficiently separated such that each track can be treated in isolation and essentially all SM processes may be easily identified. 
Using \Eq{eq::Nmu_estimation}, we estimate 
\begin{align}
    \rho 
    \sim 
    \frac{N_\mu}{ N_{\text{\tiny dev}} \SFasernu} 
    \sim 
    \frac{6.7\times 10^4}{\text{cm}^2}\left(\frac{20}{N_{\text{\tiny dev}}}\right)
    \left(\frac{\mathcal{L}}{150 \,\text{fb}^{-1}}\right)\,,
\end{align}
where $N_{\text{\tiny dev}}$ is the number of emulsion film development during the run.
We find that $r_{\text{track}}\sim 0.04~\text{mm} \ll r_{\text{\tiny M}}$. 
Therefore, identifying an excess of secondary particles could be challenging as it would have to be made on top of the pile-up generated by the surrounding tracks. 
Alternatively, one could try to identify single tracks of high-energy electrons and positrons emerging from a source muon track.
These tracks typically emerge at length scales much smaller than the Moli\`ere radius and their identification is essentially limited by the spatial resolution of the emulsion track $\sigma_{\text{\tiny pos}} \ll r_{\text{track}}$. 

Both approaches outlined above would prove to be more challenging for bremsstrahlung; any sign for the photon emission is expected to be displaced due to the propagation of the photon in matter. The probability of correctly associating a displaced energy deposition with a given track determines the bremsstrahlung BG rejection rate. Beyond the overall ability to detect excess energy deposition anywhere in the detector, the bremsstrahlung rejection rate depends on two additional factors: 
(1) the typical transverse distance travelled by the photon and 
(2) the density of relevant tracks, \ie{} tracks which could be mistakenly associated with the emitted photon. 
The former factor depends on the propagation of photons in \FASERnu{}; the mean free path of photons in tungsten is $ (9/7)X^W_0 \sim 4.5\,\mm$, and the photon angle distribution peaks around $\theta_\gamma \sim {m_\mu}/{E_\mu} \sim10^{-3}$, see \App{app:photon_study} for more details. 
A naive estimation results in an average transverse displacement of $4.5\times 10^{-3}\,\mm$ for a $100\,\GeV$ muon, which is an order of magnitude smaller than $r_{\text{track}}$. 
However, photons emitted at larger angles are exponentially more likely to propagate a larger transverse distance, therefore a reliable modeling of the photon angle distribution at large angles is necessary.
The second factor, the density of the relevant tracks, depends on the kinematic cuts and algorithms used. 
For example, the number of relevant tracks can be reduced by a factor of $\sim 10^{-2}-10^{-3}$ by considering only tracks for which a substantial energy loss was observed.
The number of relevant tracks can be further reduced by considering only the tracks for which the energy loss occurred in the same region as the track in question. 
This reduction depends on the efficiency of the algorithm used to identify the energy loss region, such as the sliding window algorithm outlined above.  
We conclude that a significant fraction of bremsstrahlung events can expected to be vetoed by applying isolation criteria. 
Further rejection could be achieved by the detection of significant EM deposits (without necessarily requiring a complete photon reconstruction) and the use of kinematic characteristics. 
The proper development and optimization of this technique requires a detailed experimental study which is beyond the scope of this work.

\section{Projected sensitivity}
\label{sec:sensitivity}

The expected luminosity per muon on a tungsten target of length $\Delta$ is given by
\begin{align}
    \frac{\mathcal{L}_{\text{eff}}}{N_{\mu}} 
=   \left(\frac{\rho_W}{m_{\rm W}}\right)\Delta
=   5.1\times10^{-15}\,  \, \text{fb}^{-1}\left( \frac{\Delta}{730\times 1.1~\text{mm}}\right)\,,
\end{align}
where we use $\rho_{\rm W} = 19.3\,\text{gr}/\text{cm}^3$,  $m_{\rm W}= 171.35\,\GeV$~\cite{Workman:2022ynf} and normalize the result to the length of \FASERnu.
The expected number of high-energy muons at \FASERnu{} with $E_\mu> 100\,\GeV$ with LHC luminosity $\mathcal{L}$ and detector surface $S$ is~\cite{FASER:2020gpr,FASER:2021mtu}
\begin{align}
    \label{eq::Nmu_estimation}
    N_\mu 
=   2\times 10^{9} \,\left(\frac{\mathcal{L}}{250 \,\text{fb}^{-1}}\right)\,
    \left(\frac{S}{25\,\cm\times 30\,\cm}\right)\,,
\end{align}
where for convenience we normalize the result to the \FASERnu{} surface area and the LHC benchmark luminosity. 
The expected muon spectrum is dominated by the low energy bins~\cite{FASER:2021mtu}. 
The energy dependence of the cross-sections and kinematic variables described in the previous sections on the initial muon energy is small; we find that the inclusion of higher energy bins only generates a small $\mathcal{O}(1\%)$ effect on the projected sensitivity. 
Thus, for our analysis it is sufficient to  assume an incoming beam of $100\,\GeV$ muons.
The total number of signal events is given by $N_{\text{\tiny sig}} = \mathcal{L}_{\text{eff}} \times \sigma (N \mu \to N \mu \,\chi)$.
The two benchmarks points we consider in this work are
\begin{align}
    &\text{\FASERnu}:\;
    \LLHC=250\,\fb^{-1}\,,\; 
    \SFasernu= 25\,\cm\times 30\,\cm\,, \;
    \DFASERnu = 730\times 1.1\,\mm \, , \nonumber
 \\
    &\text{\FASERnuTwo}\text{~\cite{Feng:2022inv}}:\;
    \LLHCHL=3\,\ab^{-1}\,,\; 
    \SFasernuTwo= (40\,\cm)^2\,, \;\DFASERnuTwo = 3300\times 2\,\mm\,.
     \nonumber
\end{align}

\begin{figure}[h]
\centering
\includegraphics[width=0.8\textwidth]{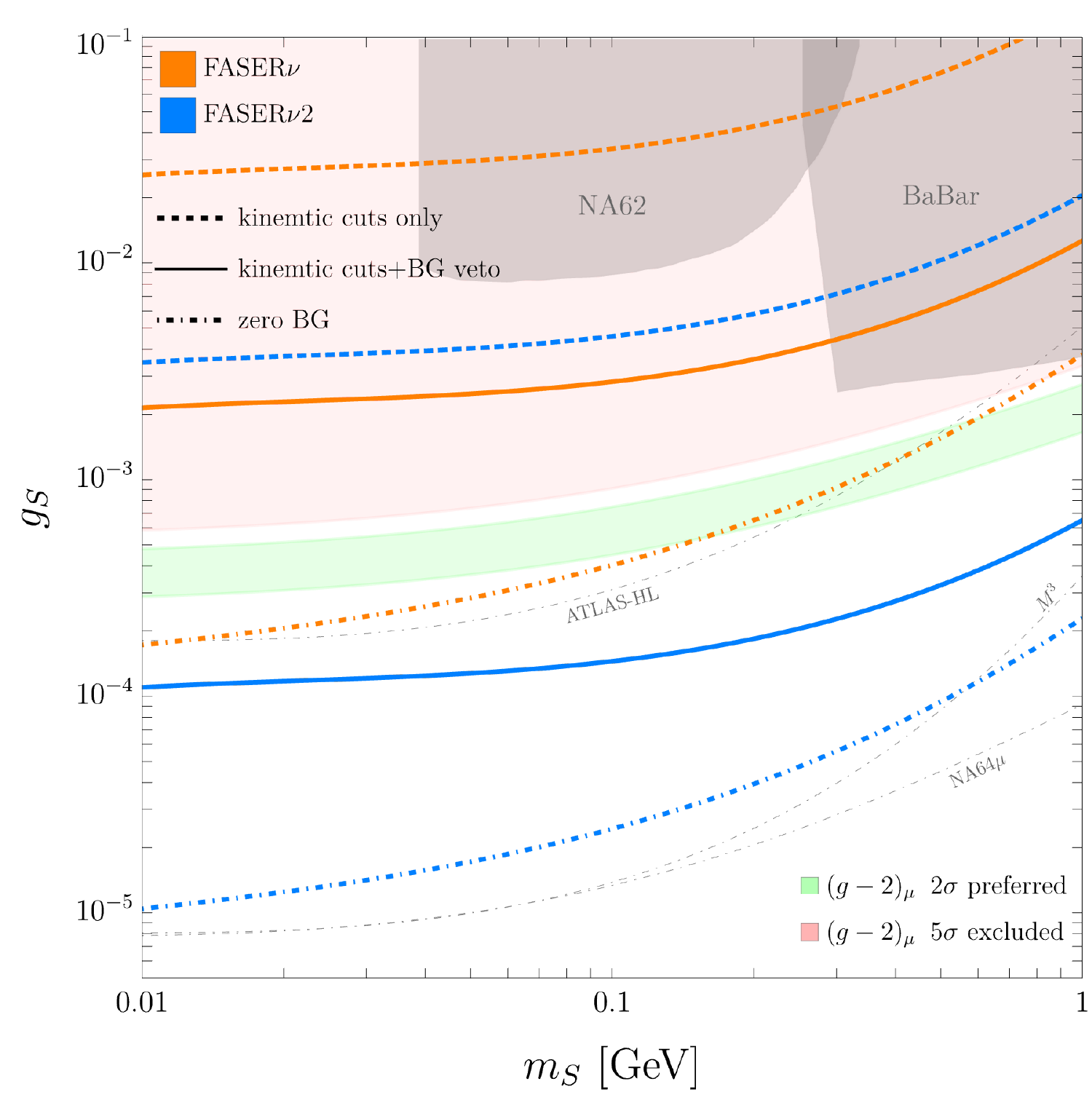}
\caption{Projected sensitivity plot for a scalar MFC for \FASERnu\,(orange) and \FASERnuTwo\,(blue). Plotted in dahsed is the projected sensitivity using only kinematic cuts and in solid using kinematic cuts and assuming BG rejection of $5\times 10^{-5}\,(10^{-6})$ for \FASERnu\,(\FASERnuTwo), see main text for more details. For reference, we plot the optimal sensitivity with $100$\% signal efficiency and zero BG events in dot-dashed. Shaded gray region is ruled out by BaBar~\cite{BaBar:2016sci,Batell:2016ove} and NA62~\cite{NA62:2021bji}.
The dot-dashed lines are the projected sensitivities from 
ATLAS-HL~\cite{Galon:2019owl}, M$^3$(phase 2)~\cite{Kahn:2018cqs} and NA64$\mu$ ($5\times 10^{12}$ muons-on-target)~\cite{Chen:2018vkr}.   }
\label{fig::scalar_sens}
\end{figure}

\begin{figure}[h]
\centering
\includegraphics[width=0.8\textwidth]{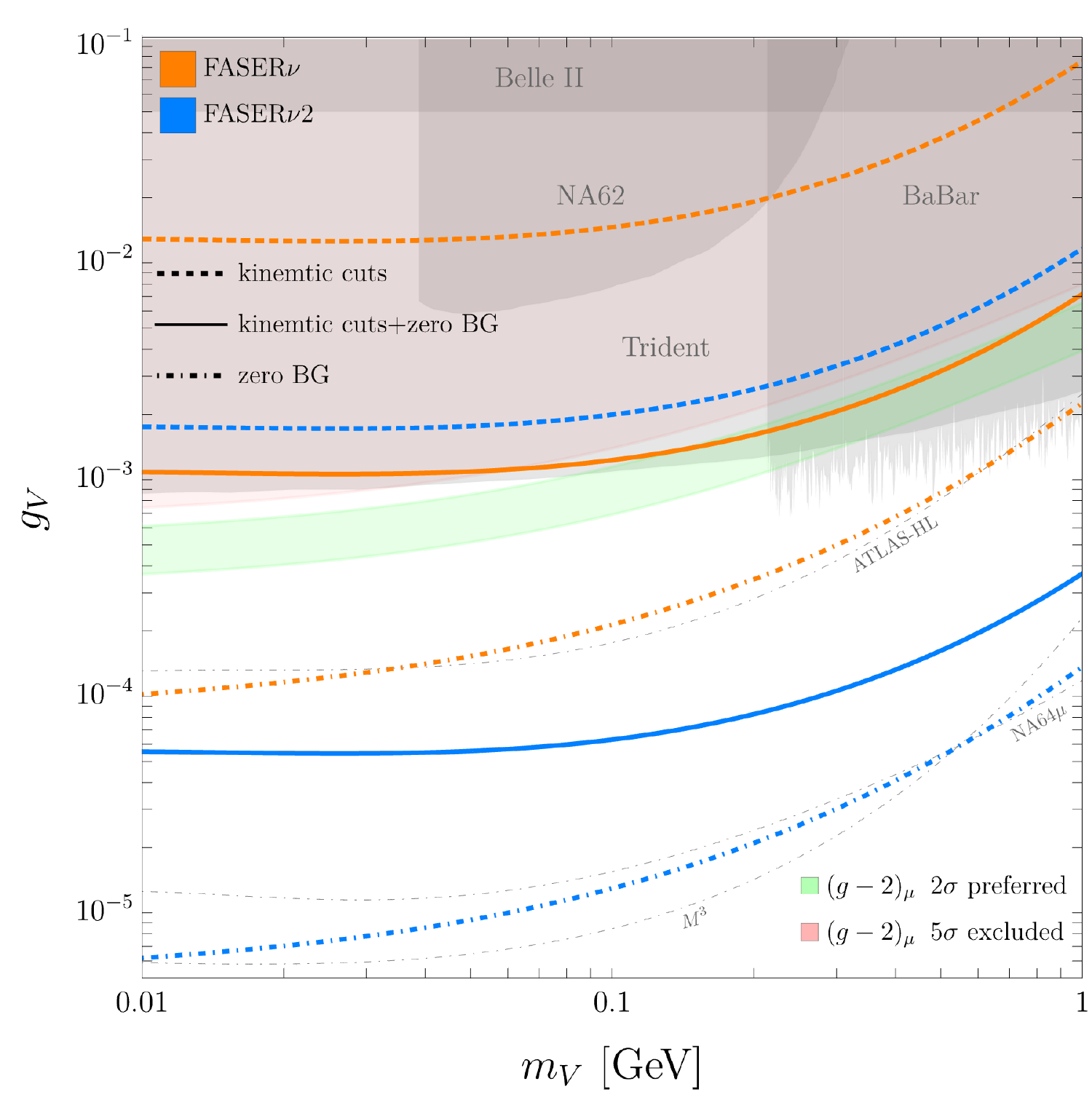}
\caption{Projected sensitivity plot for a vector MFC for \FASERnu\,(orange) and \FASERnuTwo\,(blue). Plotted in dahsed is the projected sensitivity using only kinematic cuts and in solid using kinematic cuts and assuming BG rejection of $5\times 10^{-5}\,(10^{-6})$ for \FASERnu\,(\FASERnuTwo), see main text for more details. For reference, we plot the optimal sensitivity with $100$\% signal efficiency and zero BG events in dot-dashed.
Shaded gray region is ruled out by BaBar~\cite{BaBar:2016sci}, NA62~\cite{NA62:2021bji}, Belle II~\cite{Belle-II:2019qfb} and trident production~\cite{CHARM-II:1990dvf,CCFR:1991lpl,Altmannshofer_2014}. 
The dot-dashed lines are the projected sensitivities from 
ATLAS-HL~\cite{Galon:2019owl}, M$^3$ (phase 2)~\cite{Kahn:2018cqs} and NA64$\mu$ ($5\times 10^{12}$ muons-on-target)~\cite{Gninenko:2019qiv}.   }
\label{fig::vector_sens}
\end{figure}

Before addressing BG rejection, let us first consider the optimistic zero-background case, the dot-dashed curves plotted in \Fig{fig::scalar_sens} (scalar) and \Fig{fig::vector_sens} (vector) in orange (blue) for the \FASERnu{} (\FASERnuTwo{}) benchmark case. 
In this case, \FASERnu{} probes new parameter space at low masses with couplings as small as $g_{S,V}\sim  10^{-4}$, covering the $(g-2)_\mu$ preferred region, while the reach of \FASERnuTwo{} increases by more than an order of magnitude.

However, the number of events expected from the most challenging BG source, in our case bremsstrahlung, is sizeable $B_{\text{\tiny brem}}\approx 0.07 N_\mu $. 
Therefore, achieving zero background would require an efficient BG rejection strategy. 
As a first step, we optimize the kinematic cuts on $\cR_{fi}$ and $\theta_{\mu}$ for each MFC mass point such that $S/\sqrt{B}$ is maximized. 
For the energy ratio distributions, we assume an energy resolution of $40\%$ from the MCS method. 
We introduce a signal acceptance factor $\mathcal{A}^{\text{\tiny MCS}}_{\text{\tiny S}}= 1-{(200\,\mm)}/{\Delta}$ to account for the fraction of the detector used for initial and final energy measurements. 
For the $\theta_\mu$ distribution, we use the angular resolution specified below \Eq{eq::sigma_theta}. 
We assume the position of $\theta_\mu$ within the muon track can be identified using a sliding window algorithm utilizing the MCS method, see discussion above in \Sec{sec:EnergyMeas}. 
For light masses, we find the optimal sensitivity by using looser cuts, $\theta_\mu \gtrsim 1$\,mrad and $\cR_{fi}\lesssim 0.5\,(0.7)$ for vector\,(scalar) MFC, which reduce the main BG sources, bremsstrahlung and ionization, with the respective BG rejections $\epsilon^{\text{\tiny kin.}}_{\text{\tiny brem}}\sim 0.07$ and $\epsilon^{\text{\tiny kin.}}_{\text{\tiny ion}}\sim 0.01$, while still allowing a reasonable signal efficiency $\epsilon^{\text{\tiny kin.}}_{\text{\tiny S}}\sim 0.2 $. 
For heavy masses, applying tighter cuts, $\theta_\mu \gtrsim10\,$mrad and $\cR_{fi}\lesssim 0.25\,(0.7)$ for vector\,(scalar) MFC, strongly suppresses the BG due to ionization, leaving mostly bremsstrahlung events with $\epsilon^{\text{\tiny kin.}}_{\text{\tiny brem}}\sim 0.01$, while minimally effecting the signal sensitivity which actually increases $\epsilon^{\text{\tiny kin.}}_{\text{\tiny S}}\sim 0.8 $ due to the shift of the signal to higher scattering angles and larger energy losses. 
The resulting sensitivities using the above-mentioned kinematic cuts are plotted in \Fig{fig::scalar_sens} (scalar) and \Fig{fig::vector_sens} (vector) in dashed orange and dashed blue for the \FASERnu{} and \FASERnuTwo{} benchmark cases, respectively. 
We summarize the results of our scalar and vector MFC analysis for the \FASERnu{} benchmark at the edges of the considered mass region  in \Tab{tab::scalar_vector_summary}.

The projected sensitivity achieved using only kinematic cuts represents the worst-case scenario in which none of the BG events are vetoed, leaving much room for improvement. 
The solid curves in \Figs{fig::scalar_sens}{fig::vector_sens} represent the more realistic scenario in which some of the background events that survive the kinematic cuts are vetoed due to the detection of the deposited energy in the detector.
We estimate a bremsstrahlung rejection rate of $\epsilon_{\text{\tiny brem}}^{\text{\tiny veto}}=5\times10^{-5} \,(10^{-6})$ for the \FASERnu{} (\FASERnuTwo{}) benchmark. 
While we expect higher rejection rates for the remaining BG processes, we conservatively apply the same rejection rate to all BG types.

\begin{table}[t]
\centering
\begin{tabular}{ |c|c|c|c|c|c|c|c| }
\hline
&$\cR_{fi}$ & $\theta_\mu$ & $\epsilon^{\text{\tiny kin.}}_{\text{\tiny S}}$ & $N_{\text{\tiny BG,brem}}$& $N_{\text{\tiny BG,ion}}$ & $N_{\text{\tiny BG,pair}}$& $N_{\text{\tiny BG,nuc}}$  \\ 
 \hline  \hline
  $m_S=0.01~$GeV&$\lesssim 0.7$& $\gtrsim1~$mrad & $\sim 0.2$ & $ 9.9\times 10^6$ &$8.9 \times 10^6$ &  $2.5 \times 10^5$& $2.0\times 10^6$
 \\   \hline
  $m_S=1~$GeV  & $\lesssim 0.7$ & $\gtrsim10~$mrad & $\sim 0.8$ & $1.4\times 10^6$ &$4.3\times 10^3 $&$1.1\times 10^4 $&$1.3\times 10^5$
 \\ \hline \hline
 $m_V=0.01~$GeV&$\lesssim 0.5$& $\gtrsim1~$mrad & $\sim 0.2$ & $8.5\times 10^6$ &$5.7 \times 10^6$ &  $1.7 \times 10^5$& $1.4\times 10^6$
 \\   \hline
  $m_V=1~$GeV  & $\lesssim 0.25$ & $\gtrsim10~$mrad & $\sim 0.8$ & $1.1\times 10^6$ &$4.3\times 10^3 $&$6.6\times 10^3 $&$1.3\times 10^5$
  \\
  \hline
\end{tabular}
\caption{kinematics cuts, signal efficiency $\epsilon^{\text{\tiny kin.}}_{\text{\tiny S}}$ and remaining background count for the four SM processes for our \FASERnu{} benchmark at the edges of the considered mass region. The cuts at intermediate mass points were chosen to optimize sensitivity. }
\label{tab::scalar_vector_summary}
\end{table}
\section{Outlook}
\label{sec:conclusions}

In this work, we propose to utilize  \FASERnu{} as a muon fixed target experiment to search for new muon force carriers~(MFCs), taking advantage of the large muons flux of $\sim10^9/150\,\fb^{-1}$. 
In addition to its role as a fixed target, the emulsion detector measures the muon track with excellent spatial resolution. 
This information can be used to measure kinks in the track, as well as the incoming and outgoing muon energies based on multiple coulomb scattering.  
Tracks with missing energy and large kinks are prime signatures of MFCs.

We estimate the sensitivity of the current and future runs of \FASERnu{} to the MFC parameter space.  
We find that \FASERnu{} will potentially probe previously unexplored regions of the MFC parameter space within the current run. 
Future runs will have substantially increased sensitivity and can be competitive with dedicated MFC searches such as M$^3$ and NA64$_\mu$.

This study should be considered as a proof of concept. 
Reaching the full potential of \FASERnu{} would therefore require additional dedicated studies to develop and optimize various techniques,  
\eg{} a rejection algorithm for the bremsstrahlung background events. 
Moreover, we did not include  directional information, available due to the finite width of the emulsion layers, which can potentially improve the analysis. 
Another interesting possibility would be to use rare SM processes such as muon trident production in \FASERnu{} as probe of vector MFCs. 

Interestingly, a larger muon flux is expected to be found only a few meters away from the current location of FASER. 
A well-placed dedicated muon detector could then potentially reach sensitivities well beyond the existing bounds and future dedicated searches.    

\acknowledgments
We thank Jamie Boyd and Jonathan Feng for comments on the manuscript. 
AA is supported by the European Research Council~(ERC) under the European Union’s Horizon 2020 research and innovation programme (Grant agreement No. 101002690) and JSPS KAKENHI Grant Number JP20K23373.
RB and YS are supported by grants from the NSF-BSF (No. 2021800), the ISF (No. 482/20), the BSF (No. 2020300) and by the Azrieli foundation.
IG is supported by grant from ISF (No.~751/19).
EK is supportted by grants from  NSF-BSF (No. 2020785) and the ISF (No. 1638/18 and 2323/18). 

\appendix

\section{Energy measurement analysis}
\label{app:energy_meas}

\begin{figure}[t]
\centering
\includegraphics[width=0.6\textwidth]{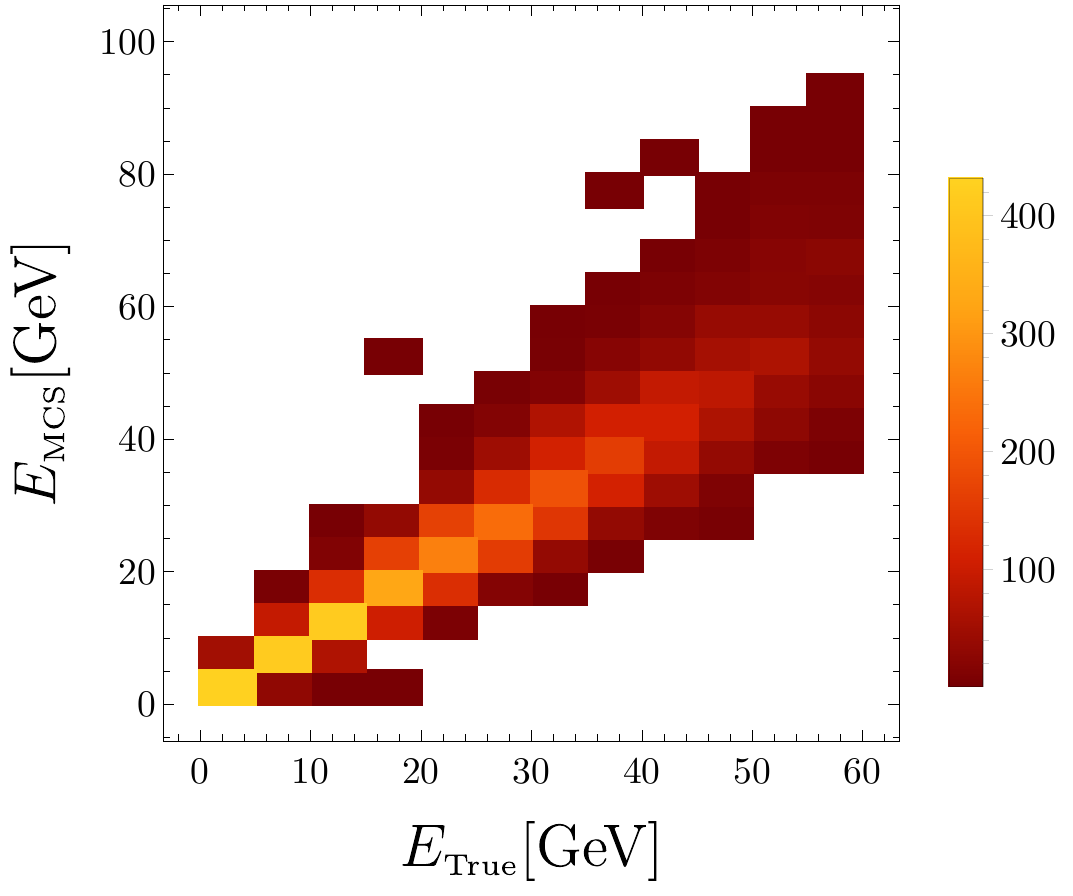}
\caption{Density distribution of $E_{\text{MCS}}$, the measured energy using the MCS method, and the true energy $E_{\text{True}}$.}
\label{fig::E_true_MCS}
\end{figure}

We performed an analysis of the energy measurement using the MCS method at lower energies $<100\,$GeV from $\cO(10^3)$ simulated tracks. 
From each track, we used 200 shifts in both the $x$ and $y$ directions to estimate the energy of the muon using the MCS method outline in \Sec{sec:EnergyMeas}.
We plot the result of our analysis in \Fig{fig::E_true_MCS}.
We find that the absolute resolution increases at lower energies due to the larger shifts compared to the spatial resolution. 
The relative resolution is approximately constant at $\approx 20\,\%$, with the exception of lower energies.
See \Tab{tab::Energy_measure} for a summary of our results. 
\begin{table}[t]

\centering
\begin{tabular}{ |c|c|c|c|c| }
\hline
 $[E_{\text{\tiny min}},E_{\text{\tiny max}}]~[\text{GeV}]$& \# of tracks & $\langle E \rangle~[\text{GeV}] $& $\sigma_E~[\text{GeV}]$ & $\sigma_E/\langle E \rangle $ \\ 
 \hline  \hline
  [0,10]& 1028 & 5.63 & 3.13&  55.6 \%
 \\   \hline
  [10,20]  & 1264  & 15.87 & 4.24  &  26.7 \%
 \\ \hline
   [20,30]  & 1214  & 25.8 & 5.36  &  20.8 \%
 \\ \hline
   [30,40]  &  1060 & 35.76 & 6.84  &  19.1\%
 \\ \hline
   [40,50]  &  726 & 45.28 &  8.4 &   18.5\%
 \\ \hline
   [50,60]  & 400  & 56.65 & 10.54  &  18.6 \%
 \\ \hline
    [99.9,100.1] & 3557 & 102.75  & 18.78 &   18.3 \%
 \\ \hline
\end{tabular}
\caption{A summary of energy measurement using the MCS method described in the text. At lower energy, larger scattering angles allow for a better (absolute) energy resolution, while the relative resolution is relatively fixed at around $20\%$ for all energies, with the exception of the first energy bin. }
\label{tab::Energy_measure}
\end{table}

\section{Bremsstrahlung in \FASERnu}
\label{app:photon_study}

High-energy photons ($\gtrsim 100\,\MeV$) are typically converted into an $e^+e^-$ pair in the nuclear electromagnetic field. 
The mean free path of the photon is given by $(9/7)X_0$, with $X_0$ the radiation length of the relevant material. 
The physical propagation distance of a high-energy photon produced inside \FASERnu, which we denote by $X_\gamma$, is distributed according to an exponential distribution  
\begin{align}
    \label{eq::brem_distance}
    \Delta_\gamma \sim \exp\left[-   \frac{X^{\text{scat}}_\gamma( X_\gamma)}{(9/7)X^W_0} \right] \,,
\end{align}
see \Eq{eq::X_scat} for the definition of $X^{\text{scat}}_\gamma( X_\gamma)$. 
In addition to the propagation distance, we expect the photon angle distribution emitted from high-energy muons to be peaked around 
\begin{align}
    \label{eq::brem_angle}
    \theta_\gamma \sim \frac{m_\mu}{E_\mu}\,.
\end{align}
This can be easily seen by examining the bremsstrahlung differential cross-section~\cite{Bethe:1934za}, where terms of the form ${\sin^2\theta}/{(E-|p|\cos\theta)^2}$ are maximized for $\cos\theta = \frac{p}{E}$. After taking the small-angle and high-energy limits, one recovers \Eq{eq::brem_angle}. 
\\

\begin{figure}[h]
\centering
\includegraphics[width=0.45\textwidth]{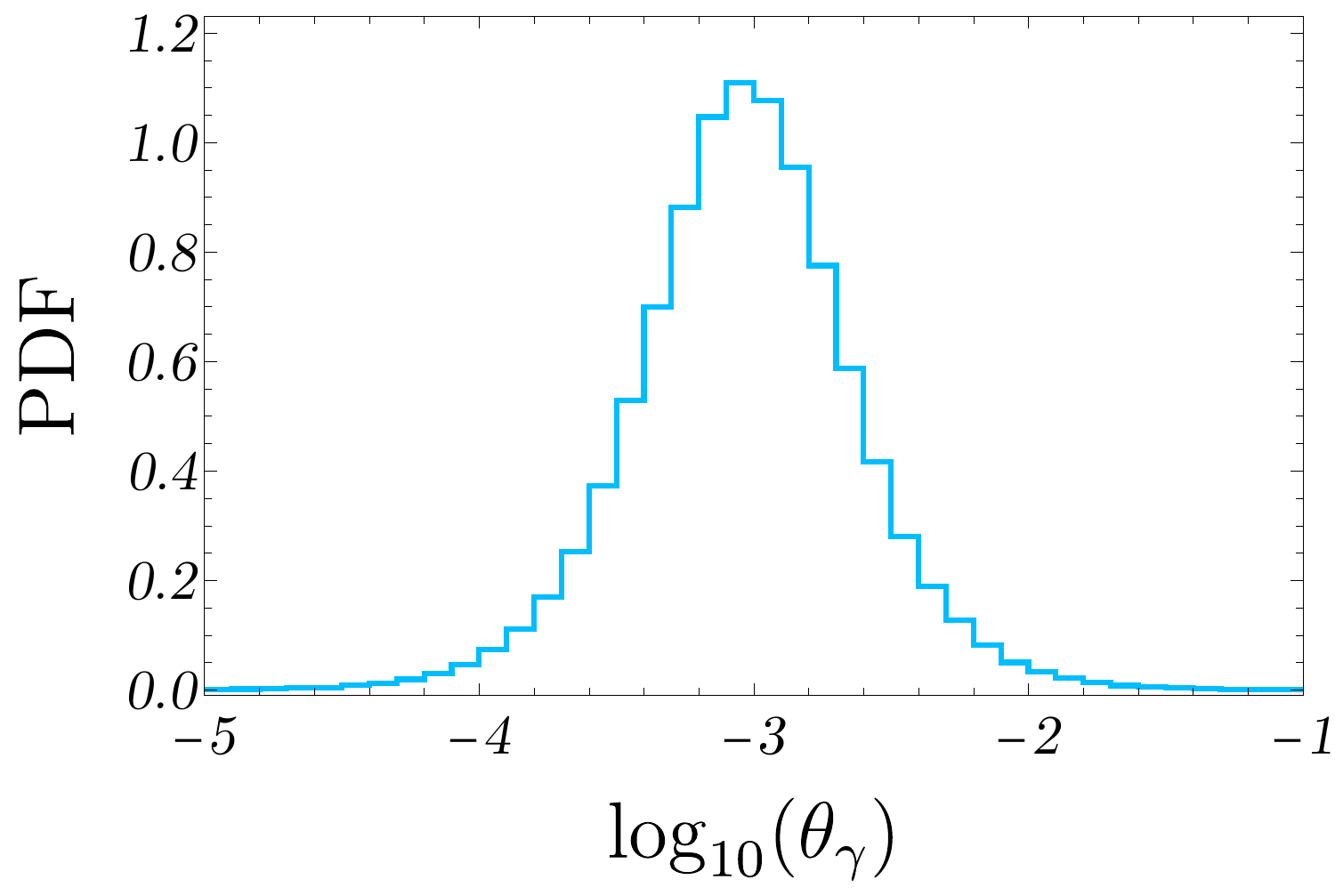}
\includegraphics[width=0.44\textwidth]{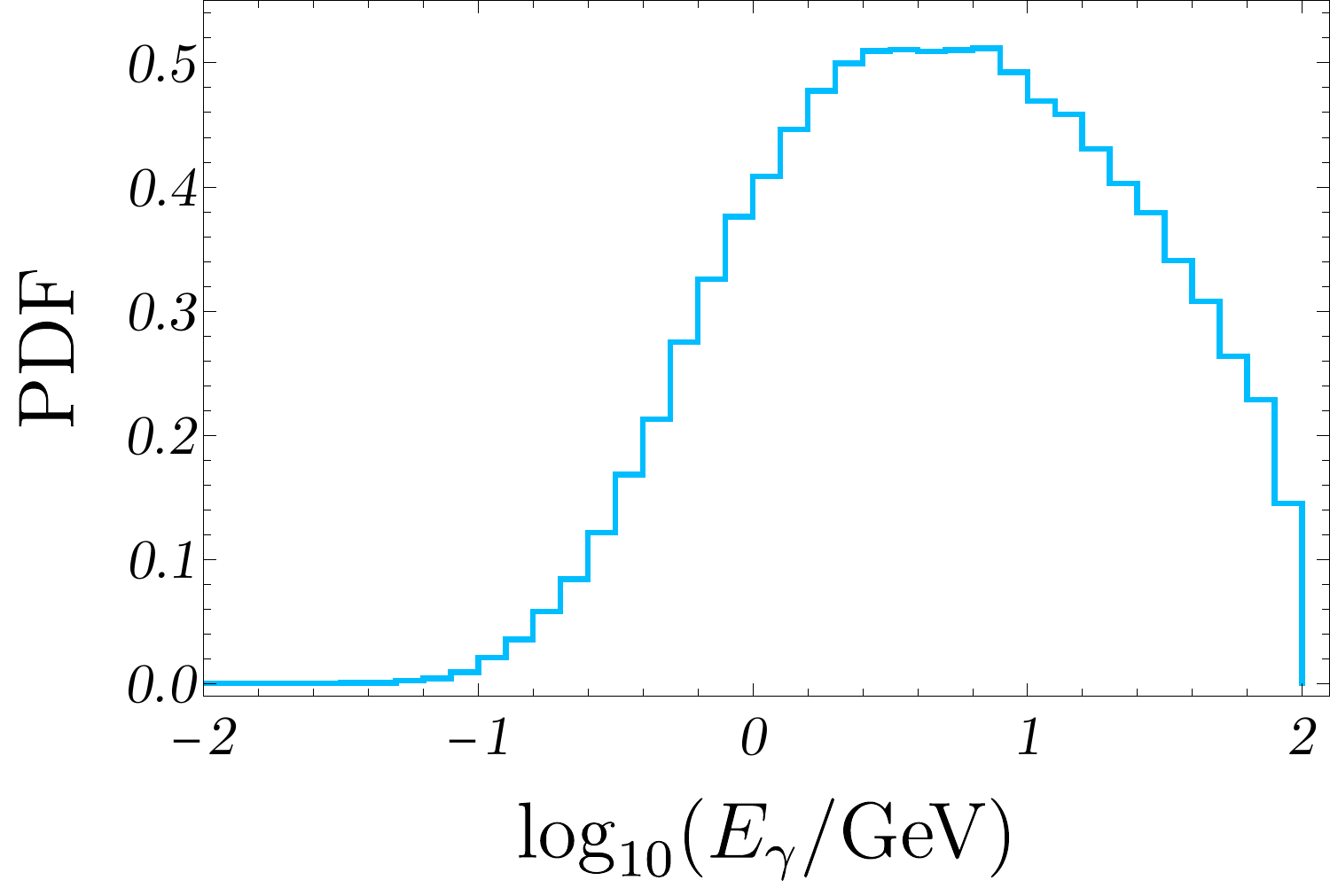}
\caption{Left: probability distribution for the emitted photon angle $\theta_\gamma$. Right: probability distribution of the emitted photon energy $E_\gamma$. }
\label{fig::brem_dist_angle_E}
\end{figure}

\begin{figure}[h]
\centering
\includegraphics[width=0.6\textwidth]{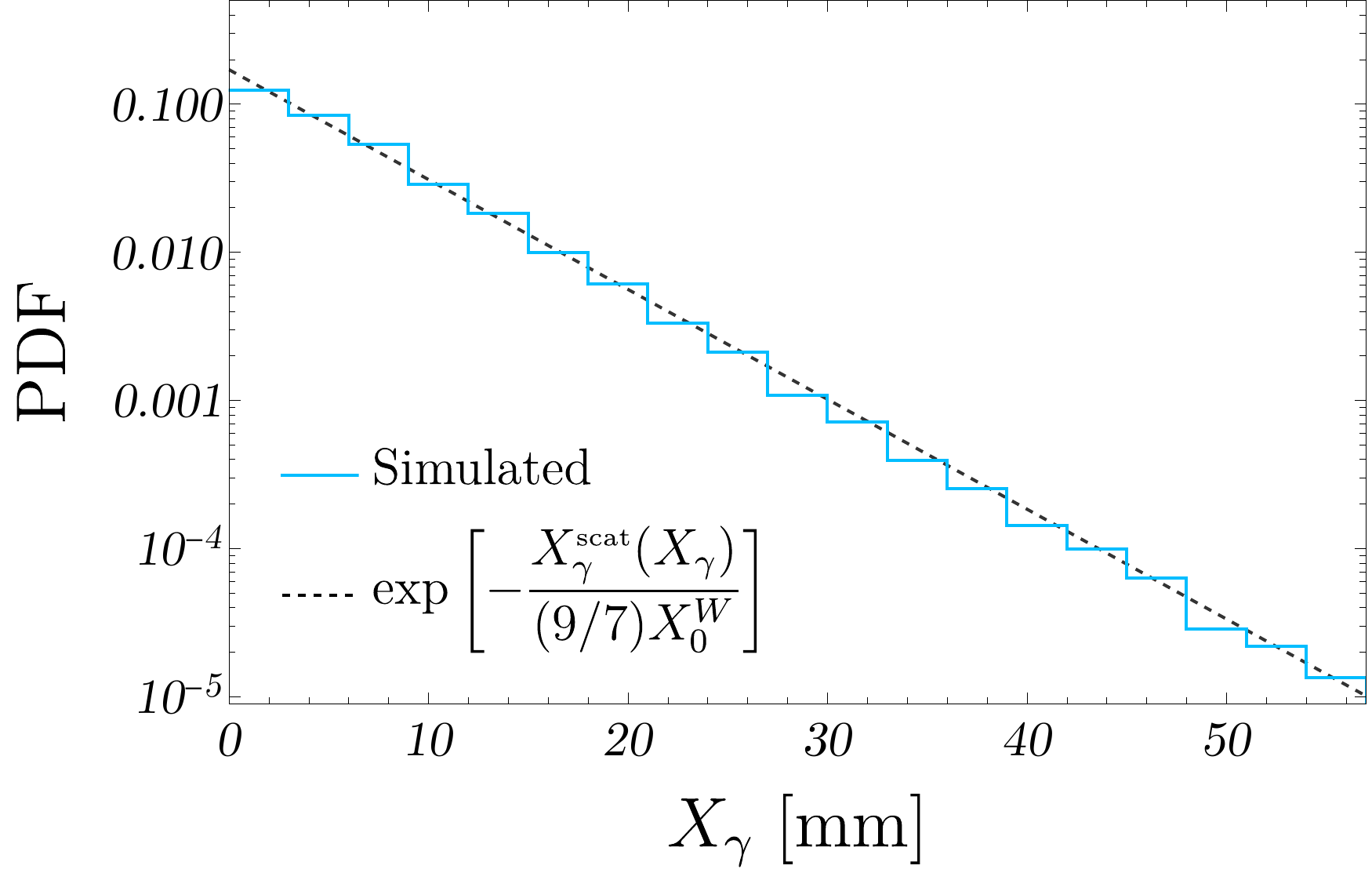}
\caption{Probability distribution of the physical propagation distance of the photon $X_\gamma$ as calculated bu simulation (solid blue) compared to the analytic expectation of \Eq{eq::brem_distance}(dashed black).}
\label{fig::brem_dist_delta}
\end{figure}

To validate \Eqs{eq::brem_distance}{eq::brem_angle}, we studied $\approx 3\times 10^5$ simulated muon tracks associated with a hard photon emission (see \Tab{tab::SM_proc}) and extracted the photon information. 
In \Fig{fig::brem_dist_angle_E}, we show the angular (left panel) and energy (right panel) distribution of the emitted photons. 
As expected, the angular distribution is strongly peaked around $m_\mu/E_\mu \sim 10^{-3}$, consistent with the simulated data used with a constant initial muon energy of $E_\mu=100\,\GeV$. 
The energy distribution is relatively flat at its bulk region, with the expected threshold at $100\,\GeV$ (the maximal available energy) and a tail for the low-energy photons. 
At shown in \Fig{fig::SM_dist}, at low energy losses,  the bremsstrahlung rate is suppressed, and therefore it is not likely that a soft $\cO(1)\,\GeV$ photon is responsible for the largest energy loss in a given track, which qualitatively explains the fast-dropping tail at low photon energies in \Fig{fig::brem_dist_angle_E}.
\\ \\
In \Fig{fig::brem_dist_delta} we show the simulated distribution of $X_\gamma$, where we find a good agreemenet with the theoretical prediction of \Eq{eq::brem_distance}. 

\bibliographystyle{JHEP.bst}
\bibliography{main.bib}

\end{document}